\documentclass[pre,twocolumn,showpacs]{revtex4}
\usepackage{amsmath}
\usepackage{amssymb}
\usepackage{amsbsy}
\usepackage{graphicx}
\usepackage{epsfig}

\def\bea{\begin{eqnarray}}
\def\eea{\end{eqnarray}}
\def\ba{\begin{eqnarray}}
\def\ea{\end{eqnarray}}
\def\beq{\begin{eqnarray}}
\def\eeq{\end{eqnarray}}
\def\be{\begin{equation}}
\def\ee{\end{equation}}

\def\bm{\begin{math}}
\def\me{\end{math}}

\begin{document}

\title{Generation of mechanical force by grafted polyelectrolytes in an electric field 
}

\author{N.~V.~Brilliantov}%
\affiliation{%
Department of Mathematics, University of Leicester, Leicester LE1 7RH, United Kingdom
}%
\author{Yu.~A.~Budkov}%
\affiliation{%
G.A. Krestov Institute of solution chemistry, Russian Academy of Sciences,
Akademicheskaya St. 1, 153045 Ivanovo, Russia
}%
\affiliation{%
National Research University Higher School of Economics, Department of Applied Mathematics, Moscow, Russia
}%
\author{C.~Seidel}%
\affiliation{Max Planck Institute of Colloids and Interfaces, Science Park Golm,
D-14424 Potsdam, Germany
}%

\affiliation{}
\date{\today}
\begin{abstract}
We study theoretically and by means of molecular
dynamics (MD) simulations the generation of mechanical force by grafted polyelectrolytes
in an external electric field, which favors its adsorption on the grafting plane. The force 
arises in deformable bodies linked to the free end of the chain. Varying the field, one
controls the length of the non-adsorbed part of the chain and hence the deformation of the target body, i.e., the arising force too. We consider target bodies with a linear force-deformation relation and with a  Hertzian one. While the first relation models a coiled Gaussian chain, the second one describes the force response of a squeezed colloidal particle. The theoretical dependencies of generated force and compression of the target body on applied field agree very well with the results of MD simulations. The analyzed phenomenon may play an important role in a future nano-machinery, e.g. it may be used to design nano-vices to fix nano-sized objects.
\end{abstract}


\maketitle

\section{Introduction}
Due to its obvious importance for applications, the response of polyelectrolytes  to
external  electric fields has been  of high scientific interest for the last few decades,
e.g., ~\cite{Muthu1987,Bajpai1997,Borisov1994,Boru98,Joanny98,Muthu2004,Dobry2000,Dobry2001,Borisov2001,
Netz2003,Netz2003a,Borisov2003,FriedsamGaubNetz2005,BrilliantovSeidel2012, SeidBudBrill2013}.                            Moreover, novel experimental techniques that allow exploration of a single polymer  chain
aided developments in this area \cite{FriedsamGaubNetz2005}.  In fact, the
ability of polyelectrolyte chains to adapt their conformation in external electric fields,
i.e., to change between expanded and contracted states when the applied field varies,
is an important property. It may be
used in future nano-machinery: possible examples of such nano-devices may be nano-vices or nano-nippers manipulated by an electric field.

\begin{figure}
\includegraphics[width=0.98\columnwidth]{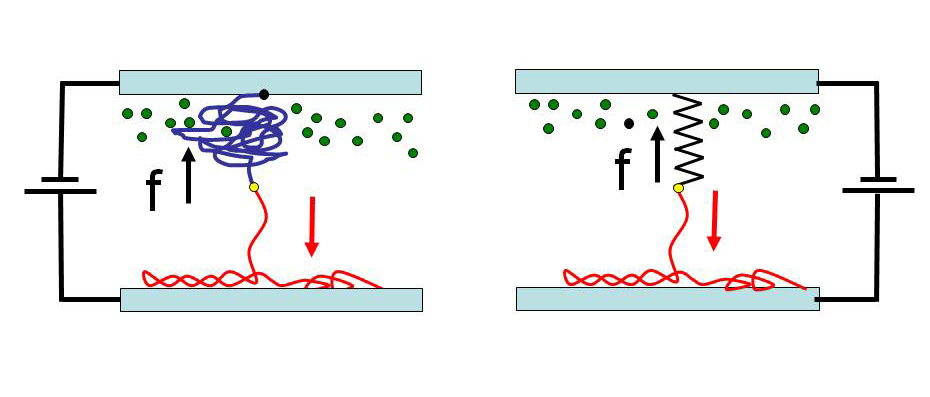}
\caption{
(Color online) Illustration of the generation of mechanical force by electric field. The electric field causes chain contraction indicated by down arrows. The restoring force $f$ of the deformed target body (up arrows) can be both linear and nonlinear, depending on the nature of the target body. The right panel shows that the target body is modelled by a spring.
}
\label{fig:spring_up}
\end{figure}

Suppose one end of a chain is fixed on a plane (i.e., the polyelectrolyte is grafted),
while the other end is linked to a nano-sized (target) body that can suffer
deformation. If the polyelectrolyte is exposed to an external electric field that favors adsorption at the grafting plane,  its conformation will be determined by both the field and the restoring force exerted by the deformed target body on the chain, see Fig. \ref{fig:spring_up}. Increased adsorption of the polyelectrolyte
in response to a changing electric field will cause a deformation of the target body
and give rise to a force acting between the chain and target. More precisely, the force will depend both on the magnitude of the deformation and the specific force-deformation
relation of the target body. Hence, by applying an electric field, one can manipulate the        conformation of polyelectrolyte chains as well as the force affecting the target body.

\begin{figure}
\includegraphics[width=0.98\columnwidth]{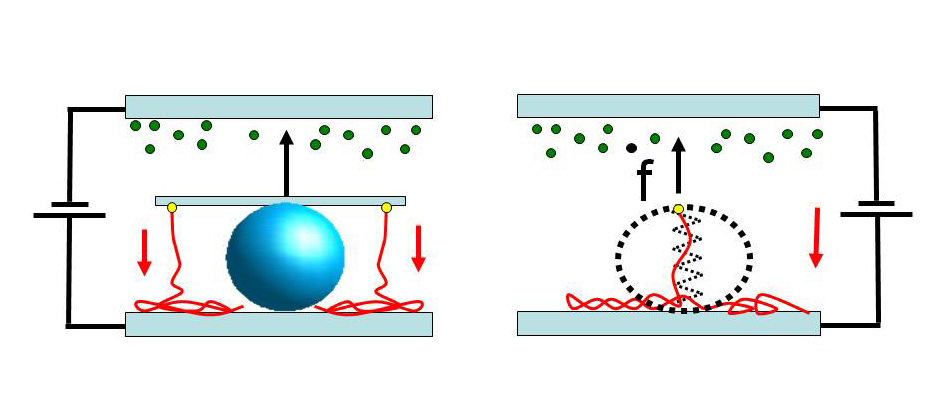}
\caption{ (Color online)
Illustration of the work principle of nano-vice: The target particle -- the colloidal particle is fixed at  sufficiently strong fields due to the polyelectrolyte chain compression; it will be released at zero field. The restoring force $f$ corresponds in this case to the Hertzian response of a compressed sphere. To illustrate a possible device, two polyelectrolyte chains are sketched, although  only one chain, linked to the Hertzian spring, was used in the simulations reported, see the right panel.
}
\label{fig:spring_down}
\end{figure}

The nature of target bodies may  be rather different, however,  the most important ones with
respect to possible applications seem to be either polymer chains or nano-particles, e.g.,
colloidal particles, see  Figs. \ref{fig:spring_up} and \ref{fig:spring_down}. In the latter
case the force-deformation relation is given by the Hertzian law, which accurately describes
the elastic response of squeezed nano-particles \cite{Hisao:2009,Hisao:2010}. On the other hand, polymer chains can exhibit  coiled states with a linear force-deformation relation or stretched conformations with a non-linear relation, e.g. \cite{GrossKhokh}.
To describe the phenomenon it is necessary to express the size of the polyelectrolyte chain as well as the force acting on the target body as a function of the applied electric field.

In the present study we address the  problem theoretically and numerically by means of molecular dynamics (MD) simulations. We analyze a model of a polyelectrolyte chain grafted to a plane, linked by its free end to a deformable target body and exposed to an external electric field. The target body is modeled by linear or non-linear springs with corresponding force-deformation  relations.  A time-independent  electric field  is applied perpendicular to the grafting plane so that it favors complete polyelectrolyte adsorption on the plane. For simplicity we consider a salt-free solution, i.e., there are only counterions that compensate the charge of the chain. For intermediate and strong electric fields (the definition is given below), additional salt leads to a renormalization of the surface charge. This happens  because the salt co-ions simply screen the plane leaving the qualitative nature of the phenomenon unchanged. Hence the salt-free case addressed here is the basic one, which allows a simpler analytical treatment. The general case of a solution with additional salt ions will be studied elsewhere~\cite{Budkov_salt}.

Counterions having  the same charge sign as the grafting plane are repelled, leaving the chain unscreened see (see Fig. \ref{fig:spring_up}). This feature is dominating if the specific volume per chain is not small and the electric field is not weak. In weak fields a noticeable fraction of counterions is located close to the chain, which leads to a partial screening of the external field and of the Coulomb interactions between monomers. Here we consider systems with a large specific volume and with fields that are not very weak. The screening of the chain in this case may be treated as a small perturbation.  We study the static case when the current across the system is zero. It is noteworthy that for the specific volumes and magnitudes of the electric field addressed here, MD simulations demonstrate a lack of the counterion screening even at finite electric current~\cite{SeidBudBrill2013}.

Here we present a first-principle theory of the phenomena and compare theoretical  predictions with MD simulations results. We observe a quantitative agreement between theory and MD data
for all magnitudes of electric field, except very weak fields when screening of the polyelectrolyte becomes significant. The simpler problem of the conformation of a grafted polyelectrolyte exposed to  a constant force in electric field, has been explored theoretically and numerically in a previous study~\cite{BrilliantovSeidel2012}. In Refs.~\cite{BrilliantovSeidel2012,SeidBudBrill2013} we also reported some simulation results for a chain linked to a deformable target body along with our previous simpler theory for the restoring force.  In the present study  we develop a first-principle theory, based on a unified approach that describes the adsorbed part of the chain as well as the bulk part under the action of the force from the target body.

The paper is organized as follows: In Section II, we present our analytical theory,
where we calculate  the free energy  of the chain and the force acting on the target body. In Section III the numerical setup is discussed and in Section IV we present the MD results and compare them with our theoretical predictions. Finally, in Section V we summarize our findings.

\section{Theory}
We consider a system, composed of a chain of $N_0+1$ monomers, which is anchored to a planar surface at $z=0$. The anchoring end-monomer is uncharged, while each of the remaining $N_0$ beads carries the charge $-qe$ ($e>0$ is the elementary charge); $N_0$ counterions of charge $+qe$ make the system neutral.
The external electric field ${\bf E}$ acts perpendicular to the plane and favors the adsorption of the chain, Fig.~\ref{fig:Setup}. The free end of the polyelectrolyte is linked to a deformable body, modeled by a spring  with various force-deformation relations. We study a few different cases. The reaction force $f$ and the energy of deformation $U_{\rm sp}$ for a linear spring reads:

\be \label{eq:1} f=-\kappa (h- h_0), \quad  \qquad U_{\rm sp}= \frac{\kappa}{2}
(h- h_0)^2.  \ee
Here $\kappa$  is the elastic constant of the spring and  $h$ and
$h_0$ are the lengths of deformed and undeformed spring, respectively. A linear force-deformation relation corresponds, for instance, to a target body given by a polymer chain in coiled Gaussian state, e.g.~\cite{GrossKhokh}. The corresponding relation for a non-linear spring has the form

\be \label{eq:2} f=\kappa \left|h- h_0 \right|^{\gamma} {\rm sign}(h_0-h),
\qquad U_{\rm sp}= \frac{\kappa}{\gamma+1}
\left|h- h_0 \right|^{\gamma+1},   \ee
where $\gamma>1$ characterizes the stiffness of the body, which may be e.g. a polymer chain in a semi-stretched conformational state, i.e., in a state intermediate between a coiled and stretched one. It is known that  stretched polymer chains demonstrate much larger stiffness than Gaussian ones~\cite{GrossKhokh}. Hence, varying  the exponent $\gamma$ one can mimic different states of a chain. From the point of view of applications it is worthwhile studying the special case of a Hertzian spring with $\gamma=3/2$, which corresponds to the elastic response of a squeezed nano-particle~\cite{Hisao:2009,Hisao:2010}, e.g., a colloidal particle:

\be
\label{eq:3} f=\kappa (h_0-h)^{3/2}\theta(h_0-h),
  \quad
U_{\rm sp}= \frac{2}{5} \kappa (h_0-h)^{5/2} \theta(h_0-h).
\ee
Here $h_0=d_c$ is the diameter of an unloaded colloidal particle and $h$ that of the deformed one. The unit Heaviside step function $\theta(x)$ reflects the fact that the Hertzian elastic respond arises for compressive deformations only. Although we performed MD simulations only for the above models of the elastic response, the theoretical analysis is given for the general case:

\be
\label{eq:Uspgen}
U_{\rm sp}=U_{\rm sp}(h- h_0) \quad  \qquad f=-\frac{\partial}{\partial h} U_{\rm sp}(h- h_0),
\ee
where again $h_0$ and $h$ are the sizes of undeformed and deformed target bodies, respectively.

To find the polyelectrolyte conformation in electric field and the force acting on the target body we evaluate the conditional free energy of the system and minimize it with respect to relevant variables. Let the number of (charged) monomers adsorbed at the (oppositely charged) plane be $N_s$, so that $N=N_0-N_s$ is the number of monomers in the bulk. Let $z_{\rm top}$ be the distance  of the "free" chain end, linked to the target body, from the charged plane and ${\bf R} $ -- the end-to-end distance of the adsorbed polymer part of $N_s$ monomers. In what follows we compute the conditional free energy $F(N, z_{\rm top}, R)$ which may be written as
 \begin{equation}
\label{eq:F_tot} F(N, z_{\rm top}, R) \approx  F_{\rm b}+F_{\rm s} + F_{\rm bs},
\end{equation}
where $F_{\rm b}= F_{\rm b}(N, z_{\rm top})$ is the free energy of the system associated with the bulk part of the chain and the target body, $F_{\rm s}=F_{\rm s}(N_s, R)$ is the free energy of the adsorbed part of the chain and $F_{\rm bs}=F_{\rm bs}(N, z_{\rm top},R)$ accounts for the interactions between the bulk and adsorbed parts.  Minimizing then $F(N, z_{\rm top}, R)$ with respect to $N$, $z_{\rm top}$ and $R$ one can find the conformation of the chain and the force acting on the target body (see the detailed discussion below).

In the present study we focus on the range of parameters where the polyelectrolyte chain is weekly screened. This allows us to treat the interaction of the chain with counterions as a small perturbation and estimate it separately; this significantly simplifies calculations. In what follows we compute separately different parts of the free energy.

\subsection{Free energy of the bulk part of the chain}
For simplicity, we use the freely jointed chain model with $b$ being the length of the inter-monomer links, that is, the size of the monomer beads. The MD simulations discussed bellow provide a justification for this model.  The location of all monomers of the chain is determined by $N_0$ vectors ${\bf b}_i={\bf r}_{i}-{\bf r}_{i+1}$, which join the centers of $i+1$-st and $i$-th monomers ($i=1,2, \ldots N_0$). It is convenient to enumerate the monomers, starting from the "free" end linked to the target body. Then the beads with numbers $1, 2, \ldots N$ refer to the bulk part of the chain and with numbers $N+1, N+2, \ldots N_0$ to the adsorbed part. The $N_0+1$-st neutral bead is anchored to the surface.   Let the centers of the adsorbed beads lie at the plane $z=0$,~\footnote{Here we ignore the off-surface loops of the adsorbed part of the chain. These may be taken into account~\cite{BrilliantovSeidel2012}, but do not give an important contribution to the total free energy for the range of parameters addressed here.} and for simplicity the anchored bead is located at the origin, ${\bf r}_{N_0+1} =0$. Then the location of the $k$-th bead of the bulk part ($k=1,2, \ldots, N)$ may be written as
\begin{eqnarray*}
   {\bf r}_{k} \!\!&=& \!\!{\bf r}_{k}- {\bf r}_{k+1} + {\bf r}_{k+1}- {\bf r}_{k+2} \ldots
   +{\bf r}_{N_0}- {\bf r}_{N_0+1} +{\bf r}_{N_0+1}\\
   \!\!&=&\!\! \sum_{s=k}^{N_0} {\bf b}_{s}=\sum_{s=k}^N {\bf b}_{s}+
\sum_{s=N+1}^{N_0} {\bf b}_{s}=
{\bf r}_{N+1}+\sum_{s=k}^N {\bf b}_{s},
\end{eqnarray*}
where  ${\bf r}_{N+1}$ is the radius vector of the $N+1$-st bead, which is a surface bead; it is  linked to the $N$-th  bead, located in the bulk.

The  inter-center distance of $i$-th and $j$-th beads reads,
\begin{equation}\label{eq:rkl}
  {\bf r}_{ij}= \sum_{s=i}^j {\bf b}_{s},
\end{equation}
where each of the vectors ${\bf b}_s$ has the same length $b$. Its orientation may be characterized by the polar $\theta_s$ and azimuthal $\psi_s$ angles, where the axis $OZ$ is directed perpendicularly to the grafting plane, Fig.~\ref{fig:Setup}. Hence, the distances between  the reference plane $z=0$ and the $k$-th bead,  as well as between the plane and the top bead are
\begin{equation}\label{eq:zk}
  z_k=b\sum_{s=k}^N \cos \theta_s, \qquad z_{\rm top}=z_1=b\sum_{s=1}^N \cos \theta_s
\end{equation}
The location of the top bead, linked to the target body, $z_{\rm top}$   determines its deformation and the elastic energy due to the body deformation,
\begin{equation}
\label{eq:Us}
U_{\rm sp}(z_{\rm top})= U_{\rm sp} ( z_{\rm top}- z_{\rm top,0}) , \quad \qquad f = -\frac{\partial U_{\rm sp}}{\partial z_{\rm top}}.
\end{equation}
Here $z_{\rm top,0}$ is the location of the top bead of the chain when the target body is not deformed. Because the chain is assumed to be weakly screened, here we ignore screening effects, which we estimate later as a perturbation. Then the potential of the external field $\varphi_{\rm ext}$  depends on $z$ simply as $\varphi_{\rm ext}(z) =-Ez$, so that the electrostatic energy of $k$-th bead, associated with this field reads $-qe\varphi_{\rm ext}(z_k)=bqeE \sum_{s=k}^N \cos \theta_s$. Hence the interaction energy of the bulk part of the chain with the external field has the form

\begin{eqnarray}
\label{eq:Ext}
H_{\rm ext} &=& \sum_{k=1}^N -qe\varphi_{\rm ext}(z_k)= bqeE \sum_{k=1}^N\sum_{s=k}^N \cos \theta_s \\ &=& bqeE\sum_{s=1}^N s\cos \theta_s.  \nonumber
\end{eqnarray}
 Now we need to take into account the electrostatic interactions between chain monomers. Because of vanishing screening we have,
\begin{equation}\label{eq:Eself}
H_{\rm self,b} = \frac12 \sum_{i=1}^N\sum_{j=1\,j\neq i}^N V( {\bf r}_i - {\bf r}_j) = \frac12 \sum_{i=1}^N\sum_{j=1\,j\neq i}^N \frac{q^2 e^2}{\varepsilon r_{ij}},
\end{equation}
where $\varepsilon$ is the dielectric permittivity of the solution. Using the Fourier transform of the Coulomb potential $V(r)=e^2q^2/\varepsilon r$, and the expression (\ref{eq:rkl}) for the inter-monomer distances, the last equation may be recast into the form (see the Appendix A):
\begin{equation}\label{eq:Eself1}
H_{\rm self,b} = \frac{q^2 e^2 }{2 \varepsilon } \sum_{s_{1}\neq
s_{2}} \int \frac{d {\bf k }}{(2\pi )^3} \left( \frac{4 \pi}{k^2} \right) e^{  i  \sum_{s=s_{1}}^{s_{2}} {\bf k}_{\perp } \cdot {\bf b}_s^{\perp } +  k_{z} \cdot  b_s^z},
\end{equation}
where ${\bf k}_{\perp }$, ${\bf b}_s^{\perp }$ and $k_{z}$, $b_s^z = b \cos \theta_s$ are respectively the transverse and longitudinal (parallel to the axis $OZ$) components of the vectors ${\bf k}$ and ${\bf b}_s$.

In the following we first compute the  partition function associated with the bulk part of the chain. We impose the condition that the distance between the surface and the top bead, attached to the target body, is $z_{\rm top}$. Then the bulk part of the partition function reads:
\begin{eqnarray}
\label{eq:Zb}
 {\cal Z}_b(z_{\rm top})  \!\!\! &=& \!\!\! \int_{0}^{2\pi}\!\!d\psi_{1} \ldots \int_{0}^{2\pi}\!\!d\psi_{N}\! \! \int_{0}^{1} \!\!d\!\cos{\theta_{1}} \ldots \int_{0}^{1}\! \! d\!\cos{\theta_{N}}
 \nonumber \\ \!\!\! &\times& \!\!\!
 e^{-\beta U_{\rm sp} -\beta H_{\rm self,b} - \beta H_{\rm ext}}
\delta\!\left(\!\!z_{\rm top} \!- \! b \sum_{s=1}^{N}\cos{\theta_{s}}\!\!\right)b, \nonumber \\
\end{eqnarray}
where $\beta =1/k_BT$, with $T$ being the temperature of the system and $k_B$ is the Boltzmann constant; the energies $U_{\rm sp}$, $H_{\rm ext}$  and  $H_{\rm self,b}$ are defined by Eqs.~(\ref{eq:Us})--(\ref{eq:Eself1}) and the factor $b$ keeps ${\cal Z}_b$ dimensionless. In Eq.~(\ref{eq:Zb}) we also assume that the vectors ${\bf b}_s$ can not be directed downwards ($\cos \theta_s \ge 0$), which guarantees that the constrain $z_s>0$, $s=1, \ldots N$ holds true; this has been confirmed in our MD simulations.

To proceed we assume that the value of $H_{\rm self,b}$ may be approximated by its average over  the angles $\psi_1, \ldots \psi_N$, that is,  $H_{\rm self,b}\approx \left<H_{\rm self,b}\right>_{\psi}$, hence  we assume that the transversal fluctuations of the polyelectrolyte chain are small.  Then, with the use of (\ref{eq:Ext}), we rewrite Eq.~(\ref{eq:Zb}) as
\begin{eqnarray}
\label{eq:Zb1}
\mathcal{Z}_b(z_{\rm top}) \!\!&=& \!\!(2\pi)^{N}\int_{0}^{1}\!\! \! d\eta_{1} \! \ldots \!\!\! \int_{0}^{1}\!\! \!d\eta_{N}  \delta \!\left(\sum_{s=1}^{N}\eta_{s}-\tilde{z}_{\rm top}\right)  \\
\!\!&\times&\!\!
\exp\!\left[\!-\beta U_{\rm sp}(z_{\rm top })\!-\! \tilde{E} \sum_{s=1}^{N}s\eta_{s}\!-\!\beta\!
\left<H_{\rm self,b}\right>_{\psi} \right], \nonumber
\end{eqnarray}
where $\eta_s=\cos \theta_s$, $\tilde{z}_{\rm top}={z}_{\rm top}/b$, $\tilde{E}=\beta q e E b$ and
\begin{eqnarray}
\label{eq:22}
\left<H_{\rm self,b}\right>_{\psi}&=&\frac{ q^2e^2}{2\varepsilon}\sum_{s_{1}\neq
s_{2}}\int\frac{d{\bf k} }{(2\pi )^3} \left( \frac{4 \pi}{k^2 } \right)  \\
& \times & \left<e^{i{\bf k}_{\perp } \cdot \sum_{s=s_{1}}^{s_{2}}{\bf b}_s^{\perp }}\right
>_{\psi}e^{ik_{z}b\sum_{s=s_{1}}^{s_{2}}\eta_{s}}. \nonumber
\end{eqnarray}
To evaluate the latter expression we exploit the following approximation,
\begin{equation}
\label{eq:etas}
\eta_s \approx \left< \eta_s \right> =\left< \cos \theta_s \right> =\frac{{z}_{\rm top }}{bN},
\end{equation}
which implies that $z_{\rm top} \leq bN$ (recall that we consider  a freely joined chain with constant links $b$) and that $\sum_{s=s_{1}}^{s_{2}}\eta_{s}\approx z_{\rm top}\left|s_{2}-s_{1}\right|/ (b\,N)$. Referring for details to Appendix A we present here the result for $H_{\rm self,b}$, averaged over transverse fluctuations:
\begin{equation}\label{eq:Hselfin}
\beta \left<H_{\rm self,b}\right>_{\psi} =\frac{l_Bq^2N^2}{z_{\rm top}}\left( \log N -1 \right),
\end{equation}
where $l_B=e^2/\varepsilon k_B T$ is the Bjerrum length.

Using the integral representation of the $\delta$-function,

$$\delta(x)=(2 \pi)^{-1} \int_{-\infty}^{+\infty} \!\!\!d \xi e^{i\xi x},$$
we recast $\mathcal{Z}_b(z_{\rm top})$ in Eq.~(\ref{eq:Zb1}) into the form
\begin{equation}
\label{eq:Zbp}
\mathcal{Z}_b(z_{\rm top})= (2\pi)^{N-1} e^{-\beta U_{\rm sp}-\beta \left<H_{\rm self,b}\right>_{\psi } } \! \! \int_{-\infty}^{+\infty} \!\!\! \! d\xi e^{-i\xi \tilde{z}_{\rm top } +W(\xi)}, 
\end{equation}
where $W(\xi)$ contains the integration over $\eta_{1}, \ldots \eta_{N}$. Its explicit expression is given in Appendix B. For large $N \gg 1$, one can use the the steepest descend method to estimate the above integral over $\xi$. Neglecting small terms we finally obtain:
\begin{eqnarray}
\mathcal{Z}_b(z_{\rm top})\approx (2\pi)^{N-1} e^{-\beta U_{\rm sp}-\beta \left<H_{\rm self,b}\right>_{\psi } -\xi_{0}\tilde{z}_{\rm top}+W(\xi_{0})}
\end{eqnarray}
where $\xi_0$ is the root of the saddle point equation, $iz_{\rm top} -\partial W /\partial \xi=0$,
\begin{equation}\label{eq:xi0sol1}
  \xi_0 \simeq \beta qe Ez_{\rm top}.
\end{equation}
and
\begin{eqnarray}
W(\xi_{0}) =  \frac{1}{\tilde{E}} \left[ {\rm Ei}(\zeta_0) - {\rm Ei}(\zeta_N) + \log\left| {\zeta_0}/{\zeta_N} \right| \right],  
\end{eqnarray}
with ${\rm Ei}(x)$ being the exponential integral function,  $\zeta_0=\xi_0-\tilde{E}$,  and $\zeta_N=\xi_0-\tilde{E}N$. (The complete expression for $\mathcal{Z}_b$ and the derivation details are given in the Appendix B).
This yields the free energy, $\overline{F}_b(z_{\rm top},N)=-k_BT \log \mathcal{Z}_b(z_{\rm top})$, associated with the bulk part of the chain without the account of counterions:
\begin{eqnarray}
\beta \overline{F}_b(z_{\rm top},N)& \approx &  \beta U_{\rm sp}(z_{\rm top} ) + \beta \left<H_{\rm self,b}\right>_{\psi } \\
&+&\xi_{0} \tilde{z}_{\rm top}- W(\xi_{0}) -  N \log 2\pi.
\nonumber
\end{eqnarray}
The impact of counterions on the conformation of the bulk part of the chain may be estimated as a weak perturbation, so that the bulk  component of free energy reads,

\begin{equation}\label{eq:Fbtot}
F_b(z_{\rm top},N) =\overline{F}_b(z_{\rm top},N) +  F_{\rm c.ch.}(z_{\rm top},N)
\end{equation}
with
\begin{equation}\label{eq:Fc_ch}
F_{\rm c.ch.}=\frac{4 \pi \sigma_c qe^2 b}{\varepsilon\tilde{E}} \frac{e^{\tilde{E}(\tilde{z}_{\rm top}-\tilde{L})}}{e^{\tilde{E} \tilde{z}_{\rm top}/N} -1} - \frac{\pi \sigma_c qe^2b}{\varepsilon}  \tilde{z}_{\rm top} N.
\end{equation}
Here $L$ ($\tilde{L} =L/b$) is the size of the system in the $OZ$-direction, $S$ is its lateral area and $e\sigma_c = eqN_0/S$ is the apparent surface charge density, associated with  the counterions. The derivation of $F_{\rm c.ch.}$ is given in Appendix C. As it may be seen from the above equation, the impact of the counterions on the chain conformation is small, provided $e\sigma_c/E \ll 1$ and $\tilde{E} \tilde{L} \gg 1$. Assuming that these conditions are fulfilled in the case of interest, the above equation simplifies to
\begin{equation}\label{eq:Fc_ch1}
\beta F_{\rm c.ch.} \simeq -\frac{z_{\rm top}}{2 \mu} N,
\end{equation}
where $\mu =1/(2 \pi \sigma_c l_B q)$ is the Gouy-Chapman length based on the apparent surface charge density $\sigma_c = qN_0/S$.

\subsection{Free energy of the adsorbed part of the chain}
Using the notations of previous section one can write the radius vector of $l$-th bead of the adsorbed part of the chain as ${\bf r}_l= \sum_{i=N_0}^l {\bf b}_i$. Then the radius vector that joins two ends of the adsorbed part reads
\begin{equation}
\label{eq:R}
{\bf R} = \sum_{i=N_0}^{N+1} {\bf b}_i = \sum_{s=1}^{N_s} {\bf d}_s,
\end{equation}
where we introduce ${\bf d}_s = {\bf b}_{N_0+1-s}$ for the sake of notation simplicity. Obviously, for the adsorbed beads we have ${\bf r}_{kl}= \sum_{s=k}^l {\bf d}_s$. Thus, the free energy of the adsorbed part may be written as
\begin{equation}
\label{eq:FZR}
\beta F_{\rm s} =- \log   \mathcal{Z}_{\rm s} (N_s, {\bf R})  ,
\end{equation}
where $\mathcal{Z}_{\rm s} (N_s,{\bf R})$ is the conditional partition function,
\begin{eqnarray}
\label{eq:Z(R)}
  \mathcal{Z}_{\rm s}(N_s,{\bf R} )&=& \int_{0}^{2\pi}d\phi_{1} \ldots \int_{0}^{2\pi}d\phi_{N_{s}}e^{-\beta H_{\rm self, s} } \nonumber \\
&\times & \delta\left(\sum_{s=1}^{N_{s}}{\bf d}_{s}-{\bf R}\right) b^2\, ,
\end{eqnarray}
where $H_{\rm self, s} = (1/2) \sum_{s_{1}\neq
s_{2}}V({\bf r}_{s_{1}}-{\bf r}_{s_{2}})$ describes self-interaction of the adsorbed monomers with the potential $V( {\bf r}_i - {\bf r}_j)$ defined in Eq.~(\ref{eq:Eself}).
The factor $b^2 $ in the above equation keeps $\mathcal{Z}_{\rm s}$ dimensionless. Since we assume that the adsorbed part of the chain forms a flat  two-dimensional structure, the integration in Eq.~(\ref{eq:Z(R)}) is performed over $N_s$ azimuthal angles $\phi_1, \ldots \phi_{N_s}$, which define the directions of $N_s$ vectors ${\bf d}_{1}, \ldots {\bf d}_{N_s}$ on the plane. Note that the evaluation of the conditional partition sum $\mathcal{Z}_{\rm s}({\bf R} )$ allows also to estimate the equilibrium configuration of the adsorbed part of the chain. Using as previously the integral representation of the $\delta$-function we recast the above equation into the form
\begin{eqnarray}
\label{eq:Z(R)_1}
&&\mathcal{Z}_{\rm s}(N_s, {\bf R} ) \!= \!\! \int \frac{d{\bf p}}{(2\pi)^2} b^2 e^{-i {\bf p} \cdot {\bf R}} \int_{0}^{2\pi}d\phi_{1} \ldots \int_{0}^{2\pi}d\phi_{N_{s}} \nonumber \\ &&~~~\times  \exp \left\{ {-\frac{\beta}{2}\sum_{s_{1}\neq
s_{2}}V({\bf r}_{s_{1}}-{\bf r}_{s_{2}})  + i {\bf p} \cdot \sum_{s=1}^{N_s} {\bf d}_s }\right\}  \\
&&~~~ = \int \frac{d{\bf p}b^2}{(2\pi)^2}  e^{-i {\bf p} \cdot {\bf R}} \mathcal{Z}_{\rm sp}({\bf p})
\left< e^{ -\frac{\beta}{2}\sum_{s_{1}\neq
s_{2}}V({\bf r}_{s_{1}}-{\bf r}_{s_{2}})} \right>_{ {\bf p}} , \nonumber
\end{eqnarray}
where we define
\begin{equation}
\label{eq:Z0q}
\mathcal{Z}_{\rm sp}({\bf p}) \!=\!\!\int_{0}^{2\pi} \!\!\!\!d\phi_{1} \!\ldots \!\!\! \int_{0}^{2\pi}\!\!\!d\phi_{N_{s}}e^{i {\bf p} \cdot\sum_{s=1}^{N_{s}} {\bf d}_s}\!\!=\!
(2\pi)^{N_{s}}\!\left[J_{0}(pb)\right]^{N_{s}}.
\end{equation}
Here $J_{0}(x)=(2\pi)^{-1}\int_{0}^{2\pi}\cos(x\cos{\phi })d\phi $ is the zero-order Bessel function; we also take into account that $({\bf p} \cdot {\bf d}_s) = p\,b \cos \phi_s$. In Eq.~(\ref{eq:Z(R)_1})  the average over the angles $\phi_{1}, \ldots \phi_{N_s}$ is denoted as
\begin{equation}
\left< (\ldots) \right>_{\bf p }=\frac{1}{\mathcal{Z}_{\rm sp}({\bf p})}\int_{0}^{2\pi}d\phi_{1} \ldots \int_{0}^{2\pi}d\phi_{N_{s}}e^{i{\bf p} \cdot\sum_{s=1}^{N_{s}}{\bf d}_{s}} (\ldots). \nonumber
\end{equation}
Referring for computational details to Appendix D, below we give the final result for the conditional partition function:
\begin{equation}
\label{eq:ZRfin}
\mathcal{Z}_{\rm s}(N_s,{\bf R} )= \frac{(2\pi)^{N_{s}}}{\pi N_{s}} \, e^{ -\frac{R^2}{N_{s}b^2}-\frac{\pi \sqrt{2}q^2 l_B N_s^2}{R} },
\end{equation}
where $R= \left| {\bf R}\right|$. From Eq.~(\ref{eq:FZR}) then follows,
\begin{eqnarray}
\label{eq:Fs}
\beta F_s (N_s, R) &=&  \frac{R^2}{N_s b^2} +\frac{\pi \sqrt{2} q^2 l_B N_s^2}{R}   \nonumber \\
 &-&N_s \log 2 \pi - \log \pi N_s .
\end{eqnarray}
Note that $N_s= N_0 -N$. If we neglect the interaction of the adsorbed part of the chain with the bulk part  we can estimate the equilibrium end-to-end distance of the adsorbed part  $R$. Minimizing $F_s(N_s, R)$ with respect to $R$ and keeping  $N_s$ fixed,  $\left( \partial F_s /\partial R \right)_{N_s}=0$,  we obtain
 the equilibrium value of $R$,
\begin{equation}
\label{eq:extR}
R= \left(q^2 b^2 l_B \pi /\sqrt{2} \right)^{1/3} N_{s} .
\end{equation}
The above equation (\ref{eq:extR}) implies that the adsorbed part is stretched,
$R \sim N_s$. Note that the condition of a stretched conformation
does not necessarily imply a linearly stretched chain. Loose configurations of
chaotic surface loops or circular conformations are also possible.

\subsection{Interaction between bulk and adsorbed parts of the chain}
The part of the free energy which accounts for interactions between the bulk part of the chain and adsorbed part  may be estimated as (see the Appendix E for more detail)
\begin{equation}
\label{eq:Fbs1}
F_{\rm bs} (N,z_{\rm top},R)  \approx  \left< H_{\rm sb} \right>_{N,z_{\rm top},R}\,.
\end{equation}
Here $H_{\rm sb}$ is the interaction energy between $N$ charged monomers of the bulk part of the chain and $N_s=N_0-N$ monomers of the adsorbed part,
\begin{equation}
\label{eq:Hsb}
\beta H_{\rm sb} = \sum_{l=1}^N \sum_{m=1}^{N_s} \frac{l_B}{|{\bf r}_l - {\bf r}_m |},
\end{equation}
where ${\bf r}_l$ is the radius vector of the $l$-th monomer of the bulk part and ${\bf r}_m$ of the $m$-th monomers of the adsorbed part and $\left< (\ldots ) \right>_{N,z_{\rm top},R}$ denotes the averaging at fixed $N$, $z_{\rm top}$ and $R$. Using the definition of vectors ${\bf b}_i$ and ${\bf d}_j$, given  in previous sections, we can write
\be
\label{eq:rlm}
{\bf r}_l - {\bf r}_m = \sum_{s=l}^N {\bf b}_{s}+ \sum_{s=1}^m {\bf d}_{s}
\ee
and recast $H_{\rm sb}$ into the form,
\begin{equation}
\label{eq:Hsb1}
\beta H_{\rm sb} \!=\! \sum_{l=1}^N \sum_{m=1}^{N_s} \!\int \!\!\frac{d {\bf k}}{(2 \pi)^3} \!\!
\left(\frac{4 \pi l_B}{ k^2} \right)e^{i {\bf k} \cdot \sum_{s=l}^N {\bf b}_{s} +i {\bf k} \cdot \sum_{s=1}^m {\bf d}_{s}} \,.
\end{equation}
In Eq.~(\ref{eq:Hsb1}) we again use the Fourier representation of the interaction potential $1/r$ given in Appendix A. Since the averaging is to be performed at fixed $N$, $z_{\rm top}$ and $R$ we can approximate the
exponential factor in (\ref{eq:Hsb1}) as
\begin{eqnarray}
\label{eq:esp_bs}
&&\left<  e^{i {\bf k} \cdot \sum_{s=l}^N {\bf b}_{s} +i {\bf k} \cdot \sum_{s=1}^m {\bf d}_{s}} \right>_{\!\!N,z_{\rm top},R}  \\
&& ~~~~~~~ \approx e^{ -\frac{k_{\perp}^2 b^2 (N-l)}{4}
\left( 1-\frac{\tilde{z}^2_{\rm top}}{N^2} \right)+ik_z\frac{z_{\rm top}}{N}(N-l) +i ({\bf k}_{\perp} \cdot {\bf R}) \frac{m}{N_s} }, \nonumber
\end{eqnarray}
where we apply the same approximations as in Eqs.~(\ref{eq:Zb1}), (\ref{eq:etas}) and (\ref{eq:Angav}) for the bulk part of the chain and a similar
one for the adsorbed part,
\begin{equation}
\label{eq:dsR}
\sum_{s=1}^m {\bf d}_{s} \approx {\bf R} (m/ N_s) .
\end{equation}
Substituting (\ref{eq:esp_bs}) into (\ref{eq:Hsb1}) and performing integration over $d{\bf k}$ (see the Appendix E for  detail) we finally obtain,
\begin{eqnarray}
\label{eq:fbsfin}
\beta F_{\rm bs} &=&  \frac{l_B N N_s}{R}\left[ \log \left(1+\sqrt{1+z^{*\,2}_{\rm top}} \right)  \right.  \\
&+& \left. \frac{1}{z^*_{\rm top}} \log \left(z^*_{\rm top} +\sqrt{1+z^{*\,2}_{\rm top}}\right)
-\log z^*_{\rm top}\right]  \nonumber \\
&-&\frac{l_B N}{R z^*_{\rm top}} \log (2 N_s z^*_{\rm top}) \nonumber
\end{eqnarray}
where $z^*_{\rm top} = z_{\rm top}/R$ characterizes the relative dimensions of the bulk and adsorbed parts of the chain.

\subsection{Dependence of the force and deformation on the external field}
Now we can determine the dependence on the electric field of the polyelectrolyte dimensions as well as the deformation of target body. Simultaneously one obtains the dependence on applied field of the force that arises between chain and target. This may be done minimizing the total free energy of the system
$$
F(N,z_{\rm top}, R)= F_{\rm b}(N,z_{\rm top})+F_{\rm s}(N_s, R)+F_{\rm bs}(N,z_{\rm top}, R)
$$
with respect to $N$, $z_{\rm top}$ and  $R$ and using $N_s=N_0-N$ and the constrain $z_{\rm top} \leq bN$ (see the discussion above). The above three components of the free energy are given respectively by  Eqs.~(\ref{eq:Fbtot}),  (\ref{eq:Fs}) and (\ref{eq:fbsfin}). This allows to find  $N$, $z_{\rm top}$ and  $R$ as functions of the applied electric field, that is, to obtain $N=N(E)$, $z_{\rm top}=z_{\rm top}(E)$ and  $R=R(E)$. Then one can compute the force acting onto the target body. It reads,
\begin{equation}\label{eq:minztop}
   \tilde{f}(\tilde{z}_{\rm top})= \tilde{E}\tilde{z}_{\rm top}
    -\frac{\tilde{l}_B q^2 N^2 (\log N -1)}{\tilde{z}_{\rm top}^2} -\frac{N}{2 \tilde{\mu}}
+\frac{\partial \beta F_{\rm bs}}{\partial \tilde{z}_{\rm top}} ,
\end{equation}
where $\tilde{f}=\beta b f(z_{\rm top})$, with $f(z_{\rm top})= - \partial U_{\rm sp} /\partial z_{\rm top}$ being  the reduced force for a particular force-deformation relation,  Eq.~(\ref{eq:Us})  and  $\tilde{\mu} = \mu/b$
 is the reduced Gouy-Chapman length. In the above equation we exploit  Eq.~(\ref{eq:xi0sol1}) for $\xi_0$ and the saddle point equation, $iz_{\rm top} -\partial W /\partial \xi=0$,  valid for $\xi=\xi_0$ (see the Appendix B).

\section{MD simulations}
%
We report MD simulations of a polyelectrolyte modeled by a freely jointed
bead-spring chain of length $N_{0}+1$. The $(N_0+1)$-th end-bead is uncharged and anchored to a planar surface at $z=0$. All the remaining $N_{0}$ beads carry one (negative)
elementary charge. Electroneutrality of the system is fulfilled by the presence of $N_{0}$ monovalent free counterions of opposite charge, i.e.,  in our simulations $q=1$ . For simplicity, we consider the counterions to have the same size as monomers. We also assume that the implicit solvent is a good one, which implies short-ranged, purely repulsive interaction between all particles, described by a shifted Lennard-Jones potential. Neighboring beads along the chain are connected by a finitely extensible, nonlinear elastic FENE potential. For the set of parameters used in our simulations, the  bond length at zero force is  $b \simeq \sigma_{LJ}$ with $\sigma_{LJ}$ being the Lennard-Jones parameter. All particles except the anchor bead are exposed to a short-ranged repulsive interaction with the grafting plane at $z=0$ and with the upper boundary at $z=L_{z}$. The charged particles interact with the bare Coulomb potential. Its strength is quantified by the Bjerrum length $l_B=e^2/\varepsilon k_B
T$. In the simulations we set $l_B=\sigma_{LJ}$ and use a Langevin thermostat to hold the temperature $k_{B}T=\epsilon_{LJ}$ with $\epsilon_{LJ}$ being  the Lennard-Jones energy parameter. For more details of the simulation model and method see Refs.~\cite{CSA00,KUM05}. The free end of the chain is linked to a deformable target body, which is modeled by springs with various force-deformation relations. In this study we considered the two cases which seem to be the most important ones in terms of possible applications: linear and Hertzian springs described by Eqs.~(\ref{eq:1}) and (\ref{eq:3}), respectively.

In the simulations we use two different setups: one where the spring is anchored at the top plane, Fig.~\ref{fig:spring_up}, right panel, while in the second setup the spring is attached to the grafting plane,  Fig.~\ref{fig:spring_down}, right panel. For simplicity, we assume that the anchor of springs is fixed and that they are aligned in the direction of the applied field, i.e., perpendicular to the grafting plane. Under this assumptions, the  instantaneous length of the spring is $L-z_{\rm top}$  in the first case and $z_{\rm top}$ in the second one, see Figs.~\ref{fig:spring_up} and \ref{fig:spring_down}.
Here we report simulation results obtained at total chain length $N_{0}=$ 320. The footprint of the simulation box is $L_{x}\times L_{y}$ = 424 $\times$ 424 (in units of
$\sigma_{\rm LJ}$) and the box height is $L_{z}=L$ = 160.

\begin{figure}
\includegraphics[width=0.90\columnwidth]{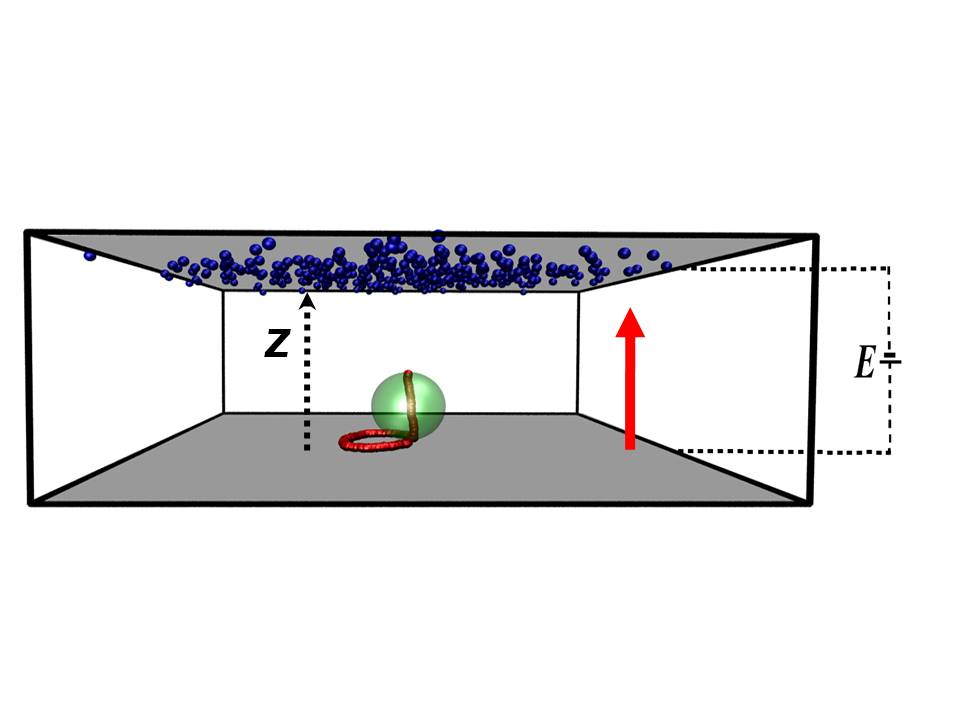}
\caption{ (Color online)
Typical simulation snapshot of a grafted polyelectrolyte exposed to electrical field $\tilde {E}=1 $, perpendicular to the grafting plane and coupled to a deformable colloidal particle of diameter $h_0= 80\,b$. The action of the particle is modeled by a Hertzian spring with spring constant $\tilde{\kappa}= \kappa b^{5/2}/k_BT=1$.  The total length of the chain is $N_0=320$. As it may be seen from the figure, for the addressed system parameters, counterions are practically decoupled from polyelectrolyte.
}
\label{fig:Setup}
\end{figure}
A typical simulation snapshot is shown in Fig.~\ref{fig:Setup}. We found,  that starting from relatively weak fields of  $Eqeb/k_BT \geq  0.1$ (recall that $qe$ is the monomer charge), the adsorbed part of the chain forms an almost flat, two-dimensional structure.  Small loops of the chain rise out of the plane up to a  height of one monomer radius. The bulk part of the polyelectrolyte is strongly stretched in perpendicular direction to the grafting plane with the inter-bead bonds being strongly aligned along the applied field. In sharp contrast to the field-free case, \cite{Winkler98,Brill98,Gole99,Diehl96,Pincus1998,MickaHolm1999,Naji:2005}  the counterion subsystem is practically  decoupled from the polyelectrolyte  which drastically simplifies the analysis.

\section{Results and discussion}
\begin{figure}
\includegraphics[width=0.99\columnwidth]{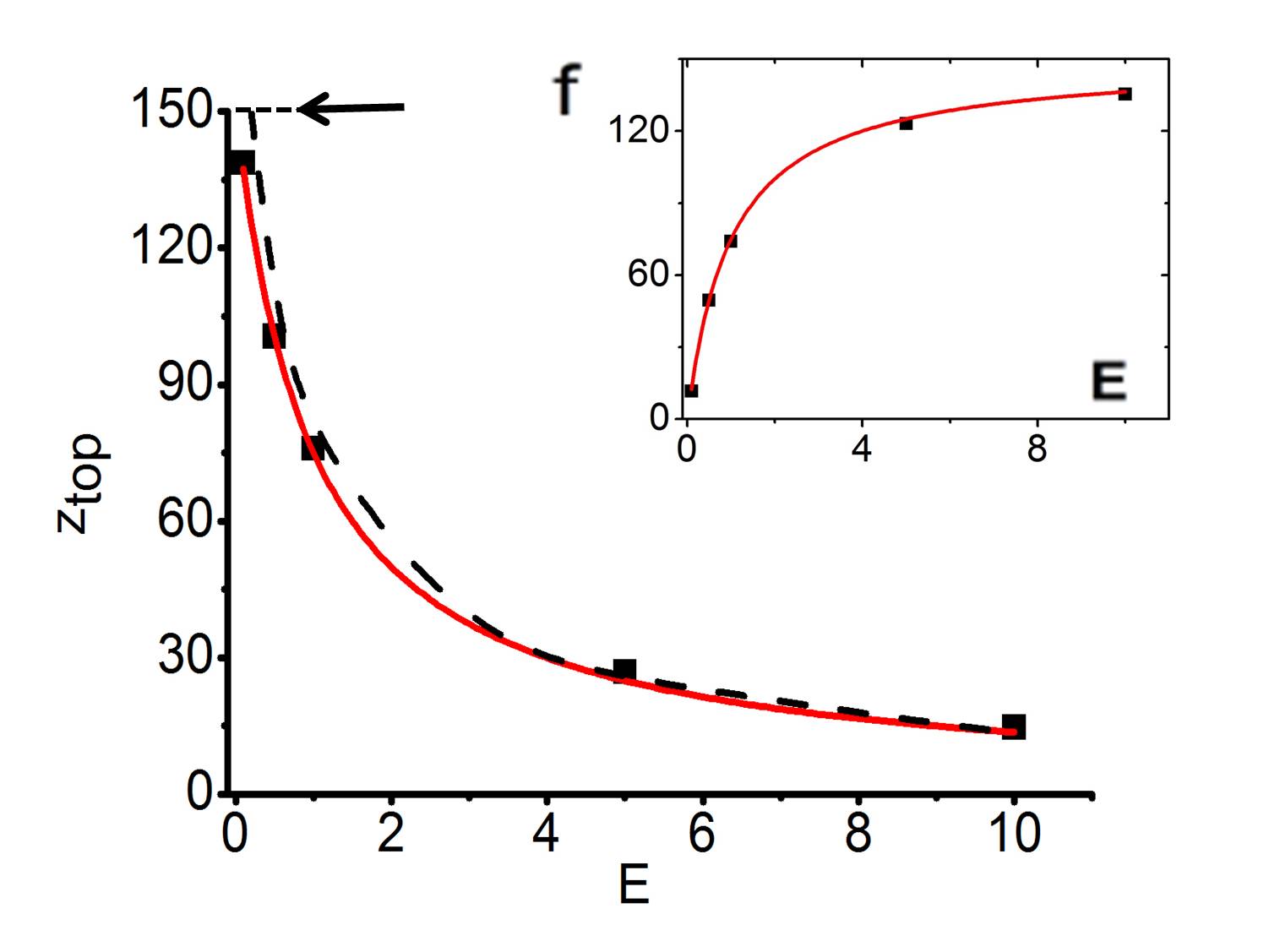}
\caption{ (Color online)
End-point height, $\tilde{z}_{\rm top} = z_{\rm top}/b$,  of a chain linked to a \emph{linear} spring, as a function of  reduced applied field $\tilde{E}=qeEb/k_BT$.
Line -- results of the theory, symbols -- MD data. The dashed  black line demonstrates $\tilde{z}_{\rm top} = N$ of the previous simplified theory \cite{SeidBudBrill2013} with $N$ taken from the MD data. Inset: reduced force generated by the applied field $\tilde{f}=fb/k_BT$ as a function of reduced field. The bar equilibrium length of the spring is $h_0=10\, b$ and its force constant is $\tilde{\kappa}= \kappa b^2/k_BT=1$. The length of the deformed spring reads, $L_z - z_{\rm top} = 160\,b -z_{\rm top}$. The total length of the chain is $N_0=320$. The arrow indicates $z_{\rm top}$ for the undeformed spring.
}
\label{fig4}
\end{figure}

\begin{figure}
\includegraphics[width=0.99\columnwidth]{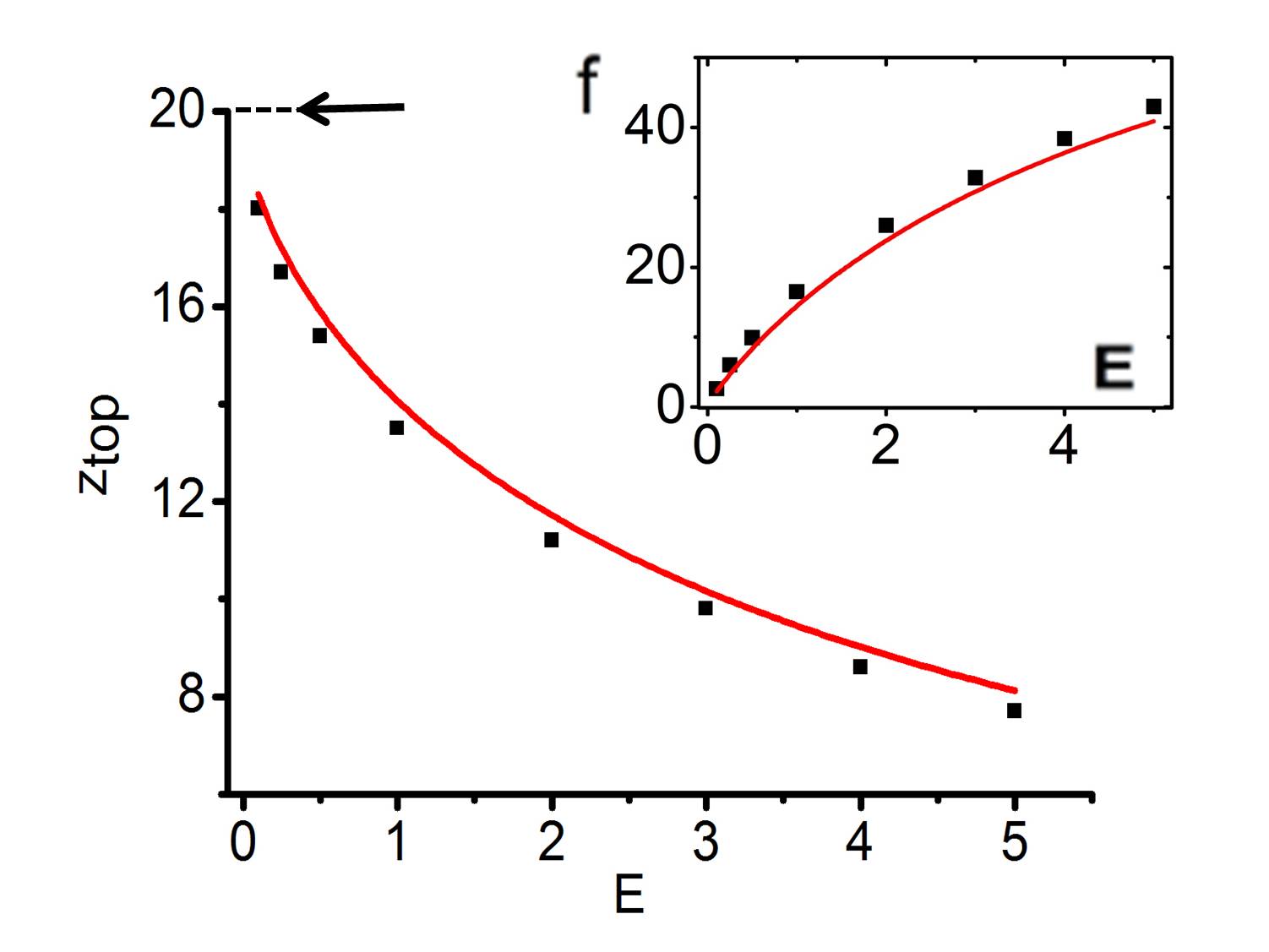}
\caption{ (Color online)
Reduced length of a Hertzian spring $\tilde{z}_{\rm} = z_{\rm top}/b$ as a function of reduced field $\tilde{E}=qeEb/k_BT$.
Line -- results of the theory, symbols -- MD data. Inset: reduced force generated by the applied field $\tilde{f}=fb/k_BT$ as a function of field. The bare equilibrium length of the Hertzian spring (undeformed colloidal particle) is $z_{\rm top,0}=d_c=20\,b$ and the force constant is $\tilde{\kappa}= \kappa b^{5/2}/k_BT=1$. The total length of the chain is $N_0=320$. The arrow indicates $z_{\rm top}$ for the undeformed spring.
}
\label{fig5}
\end{figure}

\begin{figure}
\includegraphics[width=0.99\columnwidth]{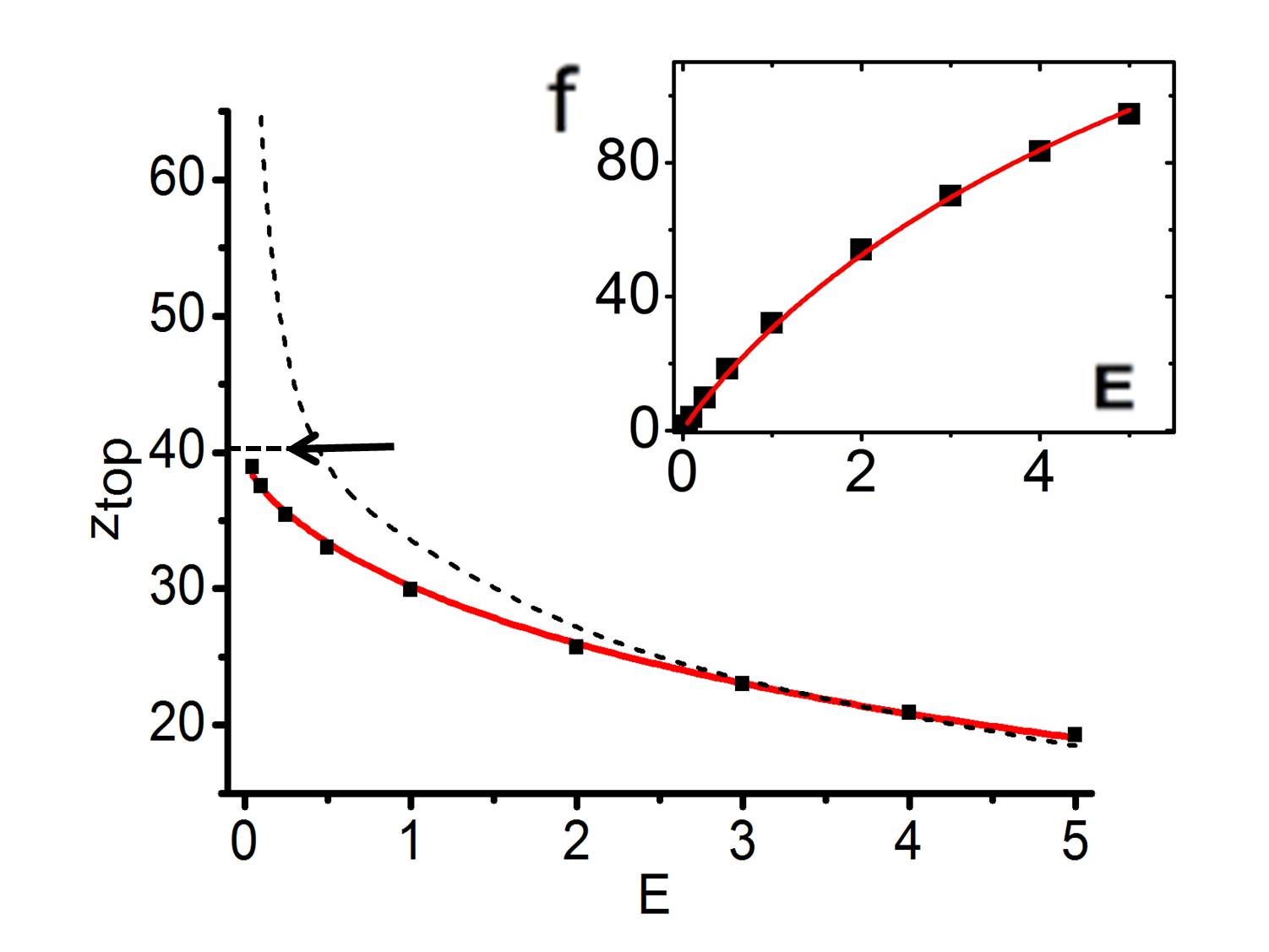}
\caption{ (Color online)
The same as Fig.\ref{fig5}, but for $z_{\rm top,0}=d_c=40\,b$. The dashed black line demonstrates $\tilde{z}_{\rm top} = N$ of the previous simplified theory \cite{SeidBudBrill2013} with $N$ taken from the MD data.}
\label{fig6}
\end{figure}

\begin{figure}
\includegraphics[width=0.99\columnwidth]{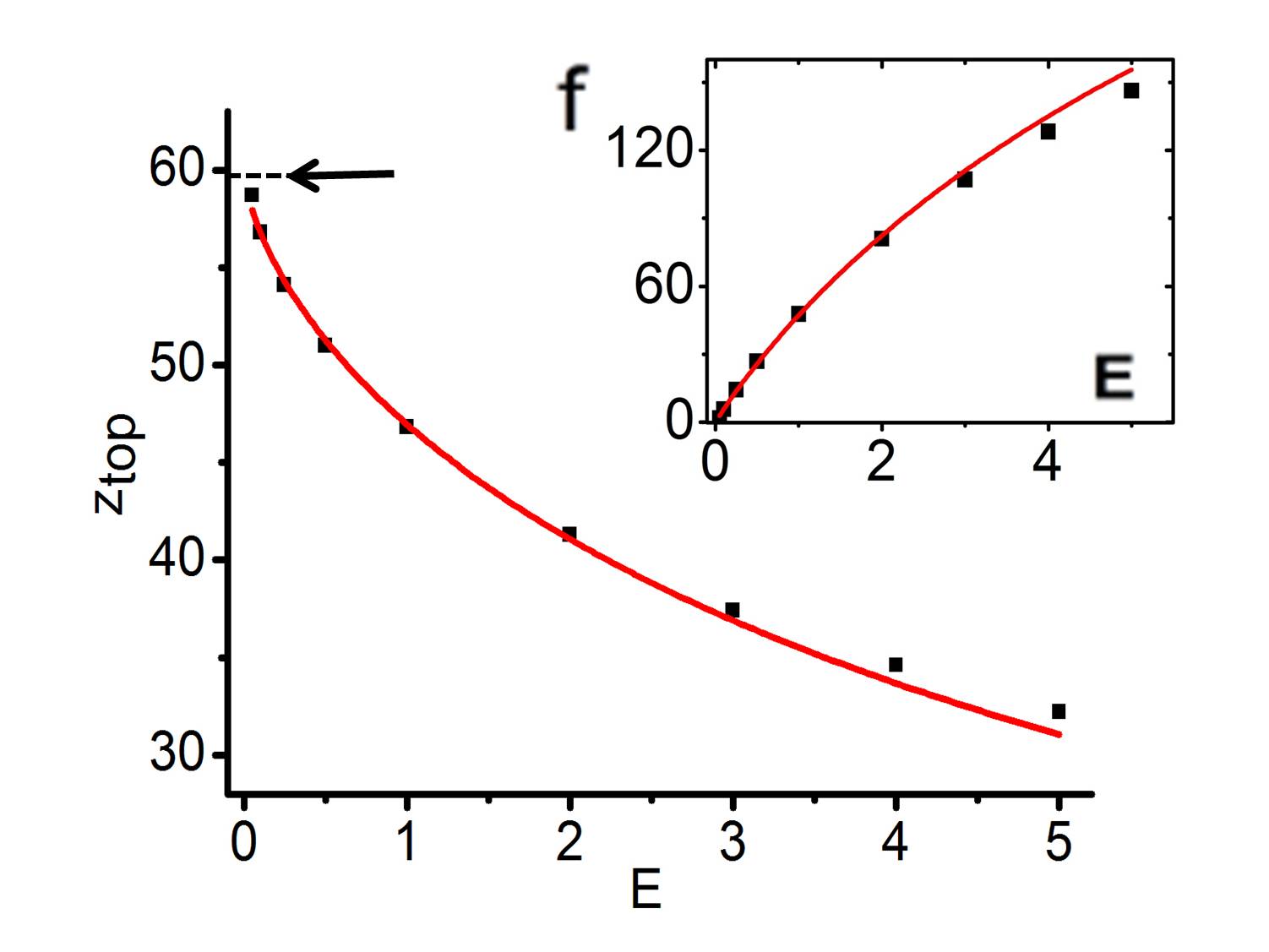}
\caption{ (Color online)
The same as Fig.\ref{fig5}, but for $z_{\rm top,0}=d_c=60\,b$.
}
\label{fig7}
\end{figure}

\begin{figure}
\includegraphics[width=0.99\columnwidth]{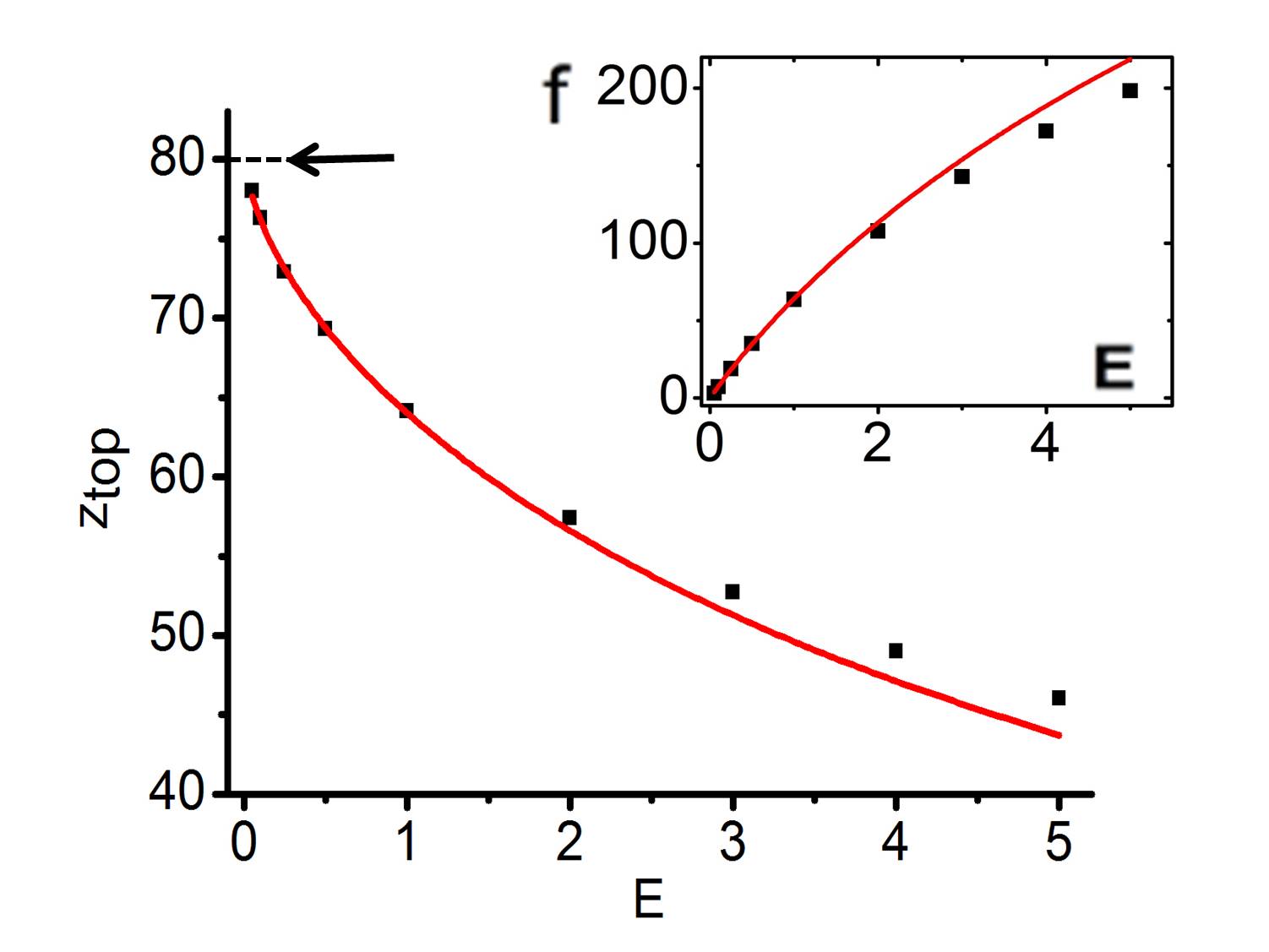}
\caption{ (Color online)
The same as Fig.\ref{fig5}, but for $z_{\rm top,0}=d_c=80\,b$.
}
\label{fig8}
\end{figure}

\begin{figure}
\includegraphics[width=0.99\columnwidth]{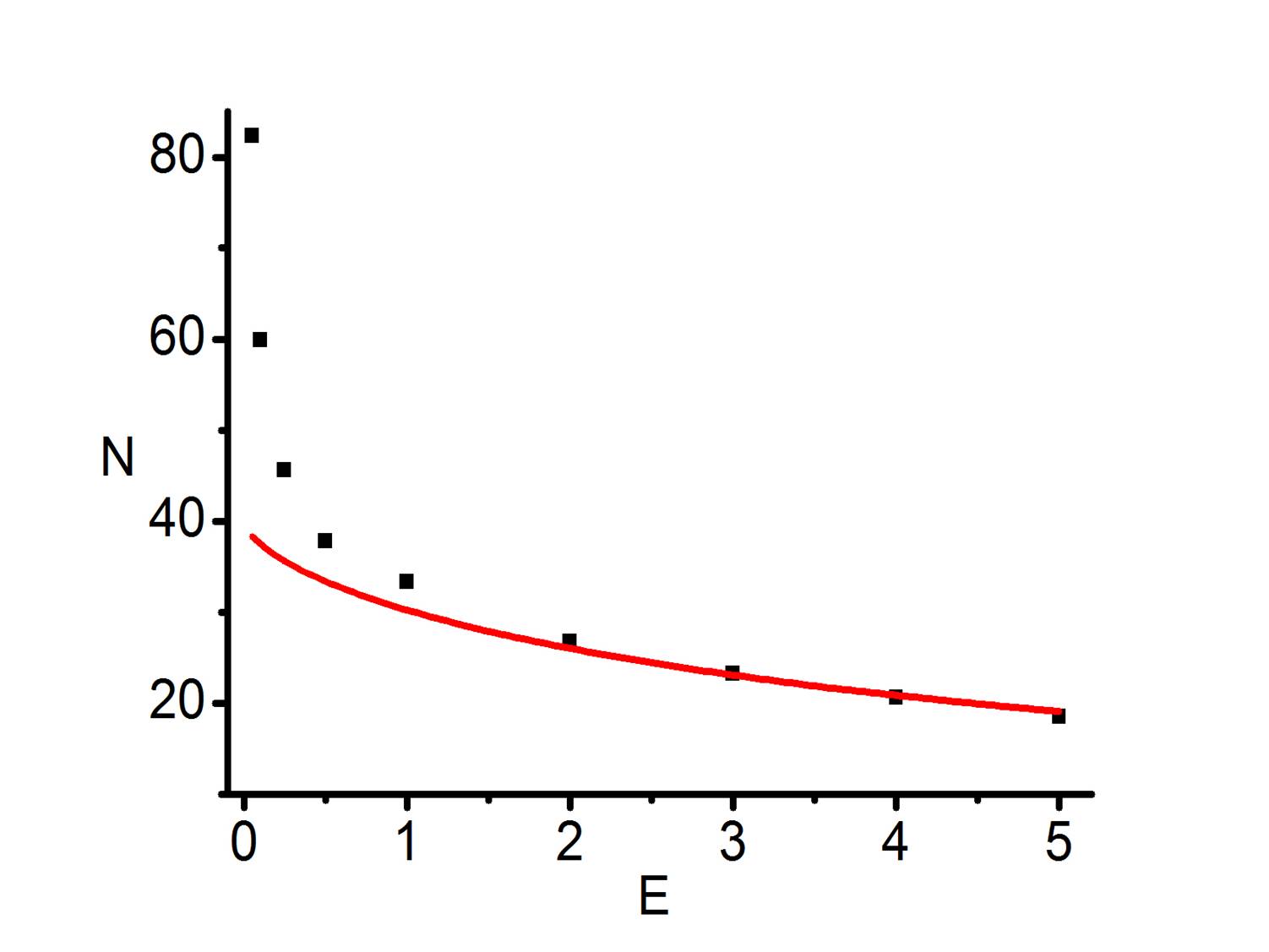}
\caption{ (Color online)
The number of chain monomers in the bulk $N$ as a function of reduced applied field, $\tilde{E}=qeEb/k_BT$.
Line -- results of the theory, symbols -- MD data. The length of the undeformed Hertzian spring is $z_{\rm top,0}=40\,b$ and the force constant is $\tilde{\kappa}= \kappa b^{5/2}/k_BT=1$. The total length of the chain is $N_0=320$.
}
\label{fig9}
\end{figure}

In Figs.~\ref{fig4} - \ref{fig8} we show  results of  MD simulations compared to the  predictions of our theory.
In particular, spring length and magnitude of the induced force are  shown as functions of the applied electric field. The spring length characterizes the deformation of the target body caused by the force acting from the polyelectrolyte chain.
Fig.~\ref{fig4} refers to a linear  spring anchored to the upper wall. Figs.~\ref{fig5} -- \ref{fig9} show the behavior of Hertzian springs of different bare equilibrium lengths (i.e. of colloidal particles of different size); these springs are anchored to the lower wall. The figures clearly demonstrate the very good agreement between theory and MD data obtained in our study.
 We wish to stress the lack of any fitting parameters used in these plots. Note however, that the theory has been developed for a highly charged chain with a relatively strong self-interaction  and interaction with the charged plane. This results in an almost flat 2D structure of the
adsorbed part of the chain and small transversal fluctuations of the bulk part;
the bond vectors of the bulk part cannot be directed down. Although the theory is rather accurate, some systematic deviations are observed for very small fields and for the shortest Hertzian springs with $z_{\rm top,0}=20\,b$.
In the latter case the deformation of the spring  and the force acting on a target body are  slightly underestimated. This possibly happens since the condition $N \gg 1$ is not as accurate for short springs as for long ones.
The theory also underestimates the number of monomer beads $N$ in a bulk for small fields. While the theory is rather accurate when $\tilde{E} >1$, there occur noticeable deviations from MD data at small fields $\tilde{E} <1$, see Fig.~\ref{fig9}.
Fortunately,  this deficiency of the new theory with respect to $N$ does not degrade the accuracy of the theoretical dependencies $z_{\rm top} (E)$ and $f(E)$, which seem to be the most important quantities in terms of possible applications. It is noteworthy that for aqueous solutions at the ambient conditions, the characteristic units of force and field are $k_BT/b \approx k_BT/l_B \approx 6\, {\rm pN}$ and $k_BT/be \approx k_BT/l_Be \approx 35\, {\rm V/\mu m}$, respectively. The latter value is about one order of magnitude smaller than the critical breakdown field for water \cite{DielBreakdown}. Another feature is worth noting.  While
the electric field alters within a relatively narrow range, the magnitude of the resulting force varies over a rather wide range, which is clearly of great interest for applications.

It is also noteworthy to compare the theoretical results of the present study with the corresponding results of the previous simplified theory, see Ref.~\cite{SeidBudBrill2013}. Some representative examples are shown in Figs.~\ref{fig4} and \ref{fig6}. Obviously the simplified theory is accurate for linear springs, except at small fields,  $\tilde{E} < 1$. At the same time it fails to satisfactorily  describe the behavior of Hertzian springs. The simplified theory drastically underestimates deformation of a  target body at small fields ($\tilde{E} < 0.5$), noticeably underestimates it at intermediate range ($1<\tilde{E} < 3$) and overestimates it in strong fields ($\tilde{E} >4$). The simplified theory has  an acceptable accuracy only in a rather narrow field interval.

 The phenomenon addressed in the present study  may be used in  future nano-machinery: A prototype of a possible nano-device, that may be called  a "nano-vice" or "nano-nippers" is illustrated in Fig.~\ref{fig:spring_down}. Here the contraction of two polyelectrolyte chains in an external electric field allows one to fix firmly a colloidal particle, which would  otherwise  perform Brownian motion. At zero or weak fields the particle will be released. Using our theory one can compute the magnitude of the field needed to keep the particle fixed, although additional knowledge about the intensity of the Brownian motion and friction forces is required. Naturally, one can think about other nano-size objects, e.g. viruses, cellular organelles or small bacteria. These objects would be characterized by other force-deformation relations.

 Consider for example nano-vices in aqueous solutions at the ambient conditions with
$l_B=0.7 \,{\rm nm}$. For simplicity we analyze the case of only one chain (see the right panel of Fig.~\ref{fig:spring_down}); the  generalization for a few chains is straightforward. Let the polyelectrolyte chain be flexible and consist of $N_0 =180$
monomers of size $b\approx l_B$, each carrying a charge $1\,e$. Let the colloidal particle
be of diameter $d_0=50b=35\,{\rm nm}$. If we use  the Young modulus $Y=0.01 {\rm GP}$,
as for rubber \cite{rubber}, for the particle material and $\nu =0.1$ for the Poisson ratio,
we obtain $\tilde{\kappa}= 2.86$~\footnote{The Hertzian force $F_H$ depends on the Young modulus $Y$, Poisson ratio $\nu$, radius of particle $R$ and deformation $\xi$ as
$F_H= \kappa \xi^{3/2}=\frac23 \frac{Y\sqrt{R}}{(1-\nu^2)} \xi^{3/2}$, see e.g. \cite{Hisao:2009,Hisao:2010}.}. In this case the field $\tilde{E} =1$ of about $35\, {\rm V/\mu m}$
generates a force of about $240 \, {\rm pN}$ and the relative deformation of $\Delta d/d_0=0.123$.  $44$ monomers remains in the bulk and $136$ are adsorbed. If the field increases up to
$\tilde{E} =2$, that is, up  to $70\, {\rm V/\mu m}$, the force increases to about $440 \, {\rm pN}$, with the deformation of $\Delta d/d_0=0.186$ and $41$ monomers in the bulk.

 Naturally, there exist plenty of other possible applications of the mechanism studied, which we plan to address in future research.

\section{Conclusion}
We analyze the generation of a mechanical force by external electric field, applied to a grafted polyelectrolyte that is linked to a deformable target body. We develop a theory of this phenomenon and perform MD simulations. The case of strong electrostatic
self-interaction of the chain and its interaction with the charged plane is addressed.
We consider target bodies with two different force-deformation relations, which seem to be the most important for possible applications: (i) a linear relation and (ii) that of a Hertzian spring. The first relation models the behavior of a coiled Gaussian chain, while the second one represents that of a squeezed colloidal particle. The theoretical dependencies of the generated force and of the compression of the target body are in a very good agreement with the simulation data. The theory, however, underestimates the number of beads $N$ of the bulk part of the chain for weak fields and small sizes of colloidal particles. Interestingly, the generated force  strongly depends on the applied electric field. While the magnitude of the  force varies over a wide interval, the field itself alters within a rather narrow range only. The  phenomenon addressed here may play an important role in  future nano-machinery. For instance, it could be utilized to design vice-like devices (nano-vices, nano-nippers) that keep nano-sized objects  fixed. Other applications of this phenomenon, which require manipulations with nano-objects, such as e.g. fusing  them together by an applied pressure are also possible.

\section{Appendix}
Here we present some calculation detail of quantities derived in the main text.
\subsection{Computation of $\left<H_{\rm self,b}\right>_{\psi}$}
First we show that $H_{\rm self,b} $ given in Eq.~(\ref{eq:Eself})  may written in the form  (\ref{eq:Eself1}). Using the integral representation of the $\delta$-function,
$$
\delta({\bf r})= (2 \pi)^{-d} \int e^{i {\bf k} \cdot{\bf r}} d{\bf k}
$$
where $d$ is the dimension of the vector ${\bf r}$, we write,

\begin{eqnarray}
\label{eq:rlm}
\frac{1}{|{\bf r}_{lm}|} &=& \frac{1}{(2\pi)^3} \int d {\bf x} \int d {\bf k}  e^{i {\bf k} \cdot
({\bf r}_{lm} - {\bf x}) }\, \frac{1}{|{\bf x}|}  \nonumber \\
&=& \frac{1}{(2\pi)^3} \int d {\bf k} \left(\frac{4 \pi}{k^2} \right) e^{i {\bf k} \cdot \sum_{s=l}^m{\bf b}_s } \\
&=&  \int \frac{d {\bf k}}{(2\pi)^3}  \left(\frac{4 \pi}{k^2} \right)  e^{i \sum_{s=l}^m {\bf k}_{\perp} \cdot {\bf b}_s^{\perp} + k_z b_s^z},  \nonumber
\end{eqnarray}
where $(4 \pi /k^2)$ is the Fourier transform of $1/{|{\bf x}|}$.  Summation of $|{\bf r}_{lm}|^{-1}$ with the prefactor $q^2e^2/2 \varepsilon$ over all $l, m =1,\ldots N$ yields Eq.~(\ref{eq:Eself1}).

To find $\left<H_{\rm self,b}\right>_{\psi}$ in Eq.~(\ref{eq:22})  first we compute the following average
\begin{equation}
\left<e^{i\sum_{s=s_{1}}^{s_{2}} {\bf k}_{\perp } \cdot {\bf b}^{\perp }_{s}}\right
>_{\psi}\!=\! \frac{1}{(2\pi)^N}\!\!\int_{0}^{2\pi}\!\! \!\!d\psi_{1} \!\ldots \int_{0}^{2\pi}\!\!\!\!\!d\psi_{\!N}e^{i\sum_{s=s_{1}}^{s_{2}}\!\! {\bf k}_{\perp} \cdot {\bf b}^{\perp}_{s}}.
\end{equation}
Due to the lateral symmetry we choose the direction of vector ${\bf k}_{\perp}$ along the $OX$ axis to
obtain,
\begin{eqnarray}
\label{eq:Angav}
&&\left <e^{i{\bf k}_{\perp } \cdot \sum_{s=s_{1}}^{s_{2}}{\bf b}^{\perp }_{s}}\right
>_{\psi} =\\ &&\quad=\frac{1}{(2\pi)^N}\int_{0}^{2\pi}d\psi_{1}..\int_{0}^{2\pi}d\psi_{N}e^{ik_{\perp}b
\sum_{s=s_{1}}^{s_{2}}\cos\psi_{s}\sin\theta_{s}} \nonumber \\
&&\quad=\prod_{s=s_{1}}^{s_{2}}\int_{0}^{2\pi}\frac{d\psi_{s}}{2\pi}e^{ik_{\perp}b\cos\psi
_{s}\sin\theta_{s}} \!= \!\!\prod_{s=s_{1}}^{s_{2}}J_{0}(k_{\perp}b\sin\theta_{s}) \nonumber \\
&&\quad=\exp \left[{\sum_{s=s_{1}}^{s_{2}}\log J_{0}(k_{\perp}b\sin\theta_{s})} \right] \nonumber \\ &&\quad \simeq
\exp \left[{\sum_{s=s_{1}}^{s_{2}}\log \left(1- k_{\perp}^2b^2\sin
^2\theta_{s}/4\right)}\right] \nonumber \\
&&\quad \approx \exp \left[ -\sum_{s=s_{1}}^{s_{2}} \frac{k_{\perp}^2b^2\sin
^2\theta_{s}}{4}\right] \nonumber \\
&&\quad \approx \exp \left[ -\frac{k_{\perp}^2b^2|s_{1}-s_{2}|}{4}\left(1-\frac{\tilde{z}_{\rm top}^2}{N^2}\right) \right], \nonumber
\end{eqnarray}
where we use the approximation $\cos ^{2}\theta_{s}\approx \tilde{z}_{\rm top}^2/N^2$  and keep in the Bessel
function expansion only the leading terms $J_0(x) =1-x^2/4 +\ldots$, where $x \sim k$. The latter approximation is justified
since the main contribution from the integrand in (\ref{eq:22}) is accumulated in the
vicinity of  $k=0$.

Using now the approximation
\begin{equation}
\label{eq:app2}
\sum_{s=s_{1}}^{s_{2}}\eta_{s}\approx \frac{\tilde{z}_{\rm top}(s_{2}-s_{1})}{N},
\end{equation}
and substituting it together with (\ref{eq:Angav}) into Eq.~(\ref{eq:22}) we obtain,
\begin{eqnarray}
\beta \left<H_{\rm self,b}\right>_{\psi } &=& \\
 \quad &=& \frac{\beta q^2e^2}{2\varepsilon}\sum_{s_{1}\neq s_{2}}\int\frac{d{\bf k }}{(2\pi
)^3}\left(\frac{4 \pi} {k_{\perp}^2+k_{z}^2} \right) \nonumber \\
 \quad &\times & e^{-\frac{k_{\perp}^2b^2|s_{1}-s_{2}|}{4}\left(1-\frac{\tilde{z}_{\rm top}^2}{N^2}\right)}
e^{i\frac{k_{z}\tilde{z}_{\rm top }|s_{1}-s_{2}|}{N}}. \nonumber
\end{eqnarray}
In the above expression, one can integrate over ${\bf k}$ (first, over $k_z$, using residues)  to  get the result

\begin{equation}\label{eq:Hselfres}
\beta \left<H_{\rm self,b}\right>_{\psi } = \frac{l_B q^2}{2}\sum_{s_{1}\neq s_{2}}\frac{\sqrt{\pi}}{2h}\,\,e^{g^2/4h^2} {\rm Erfc} \left(\frac{g}{2h} \right),
\end{equation}
where $h^2= b^2 |s_{1}-s_{2}|(1-\tilde{z}_{\rm top}^2/N^2)/4$ and $g=|s_{1}-s_{2}|\tilde{z}_{\rm top } /N$. Since $\tilde{z}_{\rm top } \sim N$ and $|s_{1}-s_{2}| \sim N \gg 1$, it is easy to show that $g/2h \gg 1$. With $e^{x^2}{\rm Erfc}(x) =(\sqrt{\pi} x)^{-1}$ for $x \gg 1$ we obtain,
\begin{eqnarray}\label{eq:Hselfres1}
\beta \left<H_{\rm self,b}\right>_{\psi } &=& \frac{l_B q^2 N}{2\, z_{\rm top}}\sum_{s_{1}\neq s_{2}}
\frac{1}{|s_{1}-s_{2}|} \\ &\simeq & \frac{l_B q^2 N}{z_{\rm top}} \int_1^{N-1}ds_1 \int_{s_1+1}^N \frac{ds_2}{s_2-s_1}  \nonumber\\
&\simeq & \frac{l_B q^2 N^2}{z_{\rm top}}(\log N -1 ),
\end{eqnarray}
that is, Eq.~(\ref{eq:Hselfin}) of the main text.

\subsection{Computation of ${\cal Z}_b(z_{\rm top})$}
From Eqs.~(\ref{eq:Zbp}) and (\ref{eq:Zb1}) follows that $W(\xi)$ is defined as,

\begin{eqnarray}
\label{eq:Wxi}
W(\xi) \!\!&=&\!\! \log \int_{0}^{1}d\eta_{1} \ldots \int_{0}^{1}d\eta_{N}\exp \left\{ \sum_{s=1}^{N}(i\xi-\tilde{E}s)\eta_{s} \right\} \nonumber \\
\!\!&=&\!\! \log \prod_{s=1}^{N}\int_{0}^{1}d\eta_{s}e^{(i\xi-\tilde{E}s)\eta_{s}} \\
\!\!&\simeq&\!\!
\int_{1}^{N}\!\!ds\left[\log(e^{i\xi-\tilde{E}s}-1)\!-\!\log(i\xi-\tilde{E}s)\right]. \nonumber \end{eqnarray}
The integral $\int_{-\infty}^{+\infty}  d\xi \exp[{-i\xi \tilde{z}_{\rm top } +W(\xi)}]$ in Eq.~(\ref{eq:Zbp}) may be estimated with
the use of the steepest descend method, that is, using the fact that for large $N$ the value of $\tilde{z}_{\rm top}$ is also large, $z_{\rm top }/b \gg 1$. Then the saddle point equation reads,
\begin{eqnarray}
\label{eq:xiztop}
&& \quad \frac{d}{d\xi}\left(-i\xi
\tilde{z}_{\rm top}+W(\xi)\right) =\\ && \qquad =-i\tilde{z}_{\rm top}+i\int_{1}^{N}ds\left[\frac{e^{i\xi-\tilde{E}s}}{e^{i\xi-\tilde{E}s}-1}
-\frac{1}{i\xi-\tilde{E}s}\right]=0. \nonumber
\end{eqnarray}
With the new variable $\xi_{0}=i\xi$, we obtain the equation that defines the implicit dependence of $\xi_{0}$ on  $\tilde{z}_{\rm top}$ and $N$:
\begin{equation}
\label{eq:xiztop1}
\tilde{z}_{\rm top} =\frac{1}{\tilde{E}}\left[\log{\frac{e^{\xi_{0}-\tilde{E}}-1}{\xi_{0}-\tilde{E}}}-
\log{\frac{e^{\xi_{0}-\tilde{E}N}-1}{\xi_{0}-\tilde{E}N}}\right].
\end{equation}
For $N \gg 1$, one can find rather accurately the solution of the above equation. Indeed, the
assumption, that $\xi_0 \sim 1 \ll N$ leads to the conclusion that $z_{\rm top} \sim \log N$, which may not hold true, neither for the coiled chain nor for the chain stretched by the force. On the other hand the assumption $\xi_0 \sim N$, which yields $\xi_0- \tilde{E} \sim N$, implies that one can apply the approximation $\log[(e^x-1)/x ] \simeq x-\log x $ at $x \gg 1$. Using the evident condition $\xi_0- \tilde{E} \gg \xi_0- \tilde{E}N$ one obtains
$$
\tilde{E}\tilde{z}_{\rm top} \simeq \xi_0 - \tilde{E} -\log (\xi_0 - \tilde{E})
$$
or
$$
\xi_0 \simeq (\tilde{z}_{\rm top}+1)\tilde{E}  + \log \tilde{z}_{\rm top}\tilde{E}.
$$
If we again take into account that $\tilde{z}_{\rm top} \sim N \gg 1$ and $\tilde{E} \sim 1 \ll N$ we arrive at an even more simple solution for $\xi_0$
$$
\xi_0 \simeq\tilde{E}\tilde{z}_{\rm top}.
$$
Hence  we obtain the following approximate expression of the partition sum,
\begin{eqnarray}
\mathcal{Z}_b(z_{\rm top})&\approx& (2\pi)^{N-1} e^{-\beta U_{\rm s}-\beta \left<H_{\rm self,b}\right>_{\psi }} \\
&\times& e^{-\xi_{0}\tilde{z}_{\rm top}+W(\xi_{0})-\frac{1}{2}\log{\frac{|W^{\prime \prime}(\xi_{0})|}{2\pi}}}, \nonumber
\end{eqnarray}
with $\xi_0$ given in the above equation and with $W(\xi_{0})$  defined by Eq.~(\ref{eq:Wxi}). It may be written as
\begin{equation}\label{eq:Wxi1}
  W(\xi_{0}) =  (1/\tilde{E}) \left[ {\rm Ei}(\zeta_0) + \log\left| \zeta_0/\zeta_N \right| - {\rm Ei}(\zeta_N) \right], 
\end{equation}
where ${\rm Ei}(x)$ is the exponential integral function and we abbreviate $\zeta_0=\xi_0-\tilde{E}$ and $\zeta_N=\xi_0-\tilde{E}N$. Similarly, we write  $W^{\prime \prime}$ as
\begin{equation}\label{eq:Wprpr}
W^{\prime \prime}(\xi_{0}) = \tilde{E}^{-1} \left[\frac{e^{\zeta_{N}}}{e^{\zeta_{N}}-1} -\frac{e^{\zeta_{0}}}{e^{\zeta_{0}}-1} -
 \frac{1}{\zeta_{N}}+\frac{1}{\zeta_{0}} \right].
\end{equation}
Finally we obtain the free energy $\overline{F}_b(z_{\rm top},N)$  associated with the bulk part of the chain (without  taking into account  counterions)
\begin{eqnarray}
&&\beta \overline{F}_b(z_{\rm top},N) \approx \beta U_{\rm sp}(z_{\rm top} ) \!+ \!\beta \left<H_{\rm self,b}\right>_{\psi } -  N \log 2\pi \nonumber\\
&&~~~~~~~~~~~~~~~+\xi_{0} z_{\rm top}- W(\xi_{0})+
\log{ |W^{\prime \prime} (\xi_{0})|^{1/2}}.
\end{eqnarray}
Note that for $N \gg 1$ the term containing $W^{\prime \prime} (\xi_{0})$ is logarithmically small as compared to other terms and may be neglected.

\subsection{Free energy of counterions}
The results of the MD simulations  show that the counterions are well separated from the chain if the field and volume of the systems are not very small. Therefore the impact of the counterions on the chain conformation may be treated as a small perturbation. Here  we perform simple estimates of the free energy of counterions. We can approximate it as,
$$
F_{\rm count} \simeq F_{\rm c.c.}+F_{\rm c.E.}+F_{\rm c.ch.},
$$
where $F_{\rm c.c.}$ is the free energy associated with the counterion-counterion interactions, $F_{\rm c.E.}$ refers to the free energy of the counterions interactions with the external field $E$ and $F_{\rm c.ch.}$ to that with the charged chain. In the case of interest one can neglect the dependence of  $F_{\rm c.c.}$ and $F_{\rm c.E.}$ on the chain conformation, so that we do not need to compute these terms. At the same time $F_{\rm c.ch.}$ can be estimated as the electrostatic energy of the chain in the additional potential $\varphi_c(z)$ caused by counterions,
\begin{equation}
\label{eq:Fcch}
 F_{\rm c.ch.} \approx \sum_{i=1}^N -qe \varphi_c(z_i).
\end{equation}

To find $\varphi_c(z)$ we start with the equilibrium Boltzmann  distribution of counterions $\rho_c(z)$ in the external field $E$ neglecting their self-interaction:
$$
\rho_c(z) = \rho_0 e^{\frac{qe E z}{k_BT}} = \frac{N_0 eq E}{S k_BT} e^{\frac{qe E (z-L)}{k_BT}},
$$
where $L$ is the size of the system in the direction along $OZ$ and $S$ is its lateral area. To obtain constant $\rho_0$
 in the above equation, we apply the normalization condition, $S\int_0^L \rho_c(z)dz =N_0$. Next we compute the electric field $E_c$ due to counterions, performing the same derivation as for the electric field of a uniformly charged plane
\begin{eqnarray}
\label{eq:Ecz}
E_c(z)\! \! &=& \! \!\frac{qe}{\varepsilon}
\int_0^L  \! \! \!dz_1 \rho_c(z_1) \int_0^{2 \pi} \! \! \! d \phi
\int_0^{\infty} \! \! \! \frac{\partial}{\partial z} \frac{r dr}{ \sqrt{(z_1-z)^2 +r^2}} \nonumber \\
\! \!&=& \! \!\frac{2 \pi eq N_0}{\varepsilon S} e^{-\frac{qe E L}{k_BT}} \left[
2e^{\frac{qe E z}{k_BT}}- e^{\frac{qe E L}{k_BT}}-1 \right] \\
\! \!&=& \! \! (4 \pi e\sigma_c/\varepsilon) e^{\tilde{E}(\tilde{z}-\tilde{L})} -(2\pi e \sigma_c/\varepsilon), \nonumber
\end{eqnarray}
where $\sigma_c = qN_0/S$ corresponds to the apparent surface charge density due to counterions and $\tilde{L}=L/b$. The second term in the above equation, $2\pi e \sigma_c/\varepsilon$,  corresponds to the renormalization of the external field $E$ due to the counterion screening of the upper plane, $E \to E - 2\pi e \sigma_c/\varepsilon$. From Eq.~(\ref{eq:Ecz}), finally we get the additional potential
\begin{equation}\label{eq:phic}
  \varphi_c(z) = 2\pi e\sigma_c z/\varepsilon - (4\pi e \sigma_c b /\varepsilon\tilde{E}) e^{\tilde{E}(\tilde{z}-\tilde{L})}.
\end{equation}
Substituting Eq.~(\ref{eq:phic}) into Eq.~(\ref{eq:Fcch})   we obtain,
\begin{equation}\label{eq:Fcch2}
 F_{\rm c.ch.} = -\sum_{i=1}^N \frac{2\pi \sigma_c}{\varepsilon}  qe^2  z_i +
  \frac{4\pi e\sigma_c b}{\varepsilon \tilde{E}}e^{-\tilde{E} \tilde{L} }  \sum_{i=1}^N e^{-\tilde{E} \tilde{z}_i }.
\end{equation}
Using $\tilde{z}_i=\sum_{s=i}^N \cos \theta_s$ (see Eq.~(\ref{eq:zk})) along with the approximation, $\cos \theta_s = \overline{\cos \theta_s}=\tilde{z}_{\rm top}/N$, we find for the first and second term in Eq.~(\ref{eq:phic}):
$$
\frac{4\pi e \sigma_c b}{\varepsilon \tilde{E}}e^{-\tilde{E} \tilde{L} }  \sum_{i=1}^N e^{-\tilde{E} \tilde{z}_i } = \frac{4 \pi e \sigma_cb}{\varepsilon\tilde{E}} \frac{e^{\tilde{E}(\tilde{z}_{\rm top}-\tilde{L})}}{e^{\tilde{E} \tilde{z}_{\rm top}/N} -1}
$$
$$
\sum_{i=1}^N \sum_{s=1}^N \frac{2\pi \sigma_c q e^2 b}{\varepsilon} \cos \theta_s =
 \frac{\pi \sigma_c q e^2 b}{\varepsilon} \, \tilde{z}_{\rm top} N\, ,
$$
which yields Eq.~(\ref{eq:Fc_ch}) of the main text.

\subsection{Computation of $ \mathcal{Z}_{\rm s}({\bf R} )$ }

We start with the computation of $\left<e^{-\beta H_{\rm self,s} }\right>_{\bf p}$.
Using only the first-order term in the cumulant expansion of the exponent we write,
\begin{equation}
\left<e^{-\frac{\beta}{2}\sum_{s_{1} \neq s_{2} }V({\bf r}_{s_{1}}-{\bf r}_{s_{2}})}\right>_{\bf p}  \approx e^{-\frac{\beta}{2}\sum_{s_{1} \neq s_{2} } \left<V({\bf r}_{s_{1}}-{\bf r}_{s_{2}})\right>_{\bf p} }.
\end{equation}
This is a mean-field approximation, which is usually a good approximation for systems with a
long-range interactions. Since $V(r)$ refers to the unscreened Coulomb interactions,
we expect this approximation to be rather accurate.

Similar as in Eq.~(\ref{eq:rlm}) we can write,
\begin{equation}
V( {\bf r}_{s_{1}}-{\bf r}_{s_{2}}) =\int \frac{d {\bf k}}{(2\pi)^3}  \tilde{V}(k) e^{i {\bf k} \cdot ({\bf r}_{s_{1}}-{\bf r}_{s_{2}})},
\end{equation}
where $\tilde{V}(k)=(q^2e^2/\varepsilon)(4\pi/k^2)$ is the Fourier transform of the interaction potential. This yields,
\begin{widetext}
\begin{eqnarray}
\label{eq:<Vr>}
 \left< V({\bf r}_{s_{1}}-{\bf r}_{s_{2}}) \right>_{\bf p}&=&
 \frac{1}{\mathcal{Z}_{0}(\bf p )} \int_{0}^{2\pi} d\phi_{1} \ldots \int_{0}^{2\pi} d\phi_{N_{s}}e^{i {\bf p } \cdot \sum_{s=1}^{N_{s}}{\bf d}_{s}} V({\bf r}_{s_{1}}-{\bf r}_{s_{2}})
=\int \frac{d {\bf k}}{(2\pi)^3}\tilde{V}(k)
\left< e^{i {\bf k} \cdot ({\bf r}_{s_{1}}-{\bf r}_{s_{2}})} \right>_{\bf p}\nonumber
\\
&=&\frac{1}{\mathcal{Z}_{0}(\bf p)} \int_{0}^{2\pi}
d\phi_{1} \ldots \int_{0}^{2\pi}d\phi_{N_{s}} \int\frac{d {\bf k}}{(2\pi)^3} \tilde{V}(k)
e^{i {\bf p } \cdot \sum_{s=1}^{N_{s}}{\bf d}_{s} + i{\bf k}_{\perp} \sum_{s_{1}}^{s_{2}}{\bf d}_{l} }\nonumber\\
\nonumber\\
&=&\int\frac{d {\bf k} }{(2\pi)^3} \tilde{V}(k)
\left[\frac{J_{0}(|{\bf k }_{\perp}+{\bf p}|\,b)}{J_{0}(pb)}\right]^{|s_{2}-s_{1}|}.
\end{eqnarray}
Here we take into account that ${\bf p}$ is a two-dimensional vector and use the definition (\ref{eq:Z0q}) of ${\mathcal{Z}_{0}(\bf p)}$. Substituting Eq.~(\ref{eq:<Vr>}) into Eq.~(\ref{eq:Z(R)_1}) we arrive at
\begin{eqnarray}
\mathcal{Z}_{\rm s}({\bf R}) &\approx&
(2\pi)^{N_{s}}\int\frac{d {\bf p} }{(2\pi)^2} e^{-i {\bf p} \cdot {\bf R}}
\left[J_{0}(pb)\right]^{N_{s}}e^{-\frac{\beta}{2}\int\frac{d{\bf k}}{(2\pi)^3}
\tilde{V}(k)\sum_{s_{1}\neq
s_{2}}^{}\left[\frac{J_{0}(|{\bf k}_{\perp}+{\bf p}|\, b)}{J_{0}(p\,b)}\right]^{|s_{1}-s_{2}|}}
\nonumber \\
&=&(2\pi)^{N_{s}}\int\frac{d{ \bf p}}{(2\pi)^2}e^{-i {\bf p} \cdot {\bf R}+N_{s}\log\left(J_{0}(pb)\right)-\frac{\beta}{2}\int\frac{d{\bf k}}{(2\pi)^3}\tilde{V}(k)\sum_{s_{1}\neq
s_{2}}^{}\left[\frac{J_{0}(|{\bf k}_{\perp}+{\bf p}|\,b)}{J_{0}(p\,b)}\right]^{|s_{1}-s_{2}|}}
\nonumber \\
&\simeq&
(2\pi)^{N_{s}}\int\frac{d {\bf p}}{(2\pi)^2}e^{ -i {\bf p} \cdot {\bf R} -\frac14 {N_{s}p^2b^2}-\frac{\beta}{2}\int\frac{d \bf k}{(2\pi)^3}\tilde{V}(k)\sum_{s_{1}\neq
s_{2}} \left[\frac{J_{0}(|{\bf k}_{\perp}+{\bf p}|b)}{J_{0}(pb)}\right]^{|s_{1}-s_{2}|}}.
\end{eqnarray}
Using the new integration variable
$${\bf G} ={\bf p} -\frac{2i {\bf R} }{N_{s}b^2},$$
we obtain
\begin{eqnarray}
\label{eq:Zs1}
\mathcal{Z}_{\rm s}({\bf R} )&=&(2\pi)^{N_s}e^{-\frac{R^2}{N_{s}b^2}}
\int\frac{d {\bf G} }{(2\pi)^2}e^{-\frac14{N_{s}b^2G^2}} \, \exp\left\{{-\frac{\beta}{2}\sum_{s_{1}\neq
s_{2}}^{}\int\frac{d{\bf k}}{(2\pi)^3}\tilde{V}(k)e^{|s_{2}-s_{1}|
\log\left[\frac{J_{0}\left(|{\bf k}_{\perp}+{\bf G}+\frac{2i{\bf R}}{N_{s}b^2}|\,b\right)}
{J_{0}\left(|{\bf G}+\frac{2i{\bf R}}{N_{s}b^2}|b\right)}\right]}}\right\}\nonumber \\
&\simeq &(2\pi)^{N_{s}}
e^{-\frac{R^2}{N_{s}b^2}}\int\frac{d{\bf G}}{(2\pi)^2}e^{-\frac14{N_{s}b^2G^2}}\exp\left\{{-\frac{\beta}{2}\sum_{s_{1}\neq
s_{2}}^{}\int\frac{d {\bf k}}{(2\pi)^3}\tilde{V}( k)e^{|s_{2}-s_{1}
|\log\left[\frac{J_{0}\left(|{\bf k}_{\perp}+\frac{2i{\bf R}}{N_{s}b^2}|b\right)}{J_{0}
\left(\frac{2i|{\bf R}|}{N_{s}b}\right)}\right]}}\right\} \nonumber \\
&\approx &(2\pi)^{N_{s}}\frac{1}{\pi N_{s}b^2}e^{-\frac{R^2}{N_{s}b^2}-\beta W_1(R)}.
\end{eqnarray}
To derive Eq.~(\ref{eq:Zs1}) we take into account that since $N_s \gg 1$,  only values of $G \sim 1/(b \sqrt{N_s})$ contribute to the above integral. The analysis also shows that $R \sim N_sb$ (see Eq.~(\ref{eq:extR})), which allows to neglect ${\bf G}$ as compared to $({\bf R}/N_s) b^2$ and to perform the Gaussian integration in the last line of (\ref{eq:Zs1}). Furthermore we define
\begin{eqnarray}
\label{eq:W1R}
\beta W_1(R)&\simeq &\frac{\beta}{2}\sum_{s_{1}\neq
s_{2}}^{}\int\frac{d{\bf k}}{(2\pi)^3}\tilde{V}(k)
e^{-\frac{i{\bf k}_{\perp}\cdot {\bf R}|s_{2}-s_{1}|}{N_{s}}}
e^{-\frac{{\bf k}_{\perp}^2b^2|s_{1}-s_{2}|}{4}} \nonumber \\
&\approx &  \frac{q^2
l_{B}}{2\pi^2}\int_{1}^{N_{s}-1}ds_{1}\int_{s_1+1}^{N_{s}}ds_{2}\int_{-\infty}^{\infty}dk_{z}\int
d{\bf k}_{\perp}\frac{e^{-\frac{i{\bf k}_{\perp}\cdot{\bf R}|s_{2}-s_{1}|}{N_{s}}}e^{-\frac{{\bf k}_{\perp}^2b^2|s_{1}-s_{2}|}{4}}}{k_{z}^2+{\bf k}_{\perp}^2},
\end{eqnarray}
where we use again the expansion of $J_0(x)$ and keep only the leading term. Integration over $k_z$ may be easily performed, yielding $\pi/k_{\perp}$. Hence we obtain,
\begin{eqnarray}
\label{eq:overkz}
&&\int_{-\infty}^{\infty}dk_{z}\int
d{\bf k}_{\perp}\frac{e^{-\frac{i{\bf k}_{\perp}\cdot{\bf R}|s_{2}-s_{1}|}{N_{s}}}e^{-\frac{{\bf k}_{\perp}^2b^2|s_{1}-s_{2}|}{4}}}{k_{z}^2+{\bf k}_{\perp}^2} = \pi \int_0^{\infty} dk_{\perp} e^{-\frac14 b^2|s_2-s_1| k^2_{\perp}} \int_0^{2 \pi} e^{-i\cos \phi k_{\perp}R|s_2-s_1|/N_s} d \phi \nonumber  \\
&&~~~~~~~= 2 \pi^2 \int_0^{\infty}  e^{-\frac14 b^2|s_2-s_1| k^2_{\perp}} J_0\left(\frac{k_{\perp}R|s_2-s_1|}{N_s} \right) dk_{\perp}
 =\pi^{5/2} \frac{e^{-\frac{R^2|s_2-s_1|}{2 N_s^2 b^2}}}{b \sqrt{|s_2-s_2|}} \, {\rm I}_0 \left( \frac{R^2|s_2-s_1|}{2 N_s^2 b^2}\right)
\end{eqnarray}
where ${\rm I}_0(x)$ is the modified Bessel function of the first kind. Substituting the above result into Eq.~(\ref{eq:W1R}) we observe that since $R\sim bN_s$, the main contribution in the integrals over $s_1$ and $s_2$ comes from the region where $|s_2-s_1|$ is small; here we can approximate  ${\rm I}_0(x)\approx 1$~\footnote{More precisely, the function $e^{-x^2} {\rm I}_0(x)$ is rather close to $e^{-x^2}$, when $x = R^2 |s_2-s_1|/N_s^2b^2$ is of the order of unity; this guarantees that the discussed approximation has an acceptable accuracy.}.  Therefore we can write,
\begin{equation}
\beta W_1(R) \approx
q^2 \tilde{l}_B\sqrt{\pi}N_s^{3/2} \int_0^1 dx \int_x^1 dy \, \frac{e^{-\frac{R^2}{2N_sb^2} |y-x|}}{\sqrt{|y-x|}}
=q^2 \tilde{l}_B\sqrt{\pi}N_s^{3/2} H\left(\frac{R^2}{2 N_s b^2} \right).
\end{equation}
\end{widetext}
Here the function $H(x)$ reads;
$$
H(x)=\frac{\sqrt{\pi}erf(\sqrt{x})(x-1/2)+xe^{-x}}{x^{3/2}},
$$
it behaves as $H(x) \simeq \sqrt{\frac{\pi}{x}}$ for $x \gg 1$. Hence, for $R^2 \gg N_s b^2$ we obtain,
\begin{equation}
\beta W_1(R) = \frac{\pi \sqrt{2} q^2l_B N_s^2}{R}.
\end{equation}
and finally, the conditional partition function,
\begin{eqnarray}
\label{eq:Zs2}
\mathcal{Z}_{\rm s}({\bf R} )\simeq \frac{{\,\, \,}(2\pi)^{N_{s}}}{\pi N_{s}}\, e^{-\frac{R^2}{N_{s}b^2}- \frac{\pi  \sqrt{2} q^2 l_B N_s^2}{R} }.
\end{eqnarray}

\subsection{Calculation  of $F_{bs}(N, z_{\rm top}, R)$}
The conditional free energy of the system  $F(N, z_{\rm top}, R)$ may be written in the following form:
\begin{widetext}
\begin{eqnarray}
\label{eq:FNZR1}
e^{-\beta F(N,z_{\rm top}, R)} &=&\int_0^{2 \pi} d\psi_1\ldots  d\psi_N \int_0^{1} d \cos \theta_1 \ldots \int_0^{1} d \cos \theta_N \delta \left(z_{\rm top} - b \sum_{s=1}^N \cos \theta_s \right) b\\
&\times& \int_0^{2 \pi} d\phi_1\ldots  d\phi_{N_s} \delta \left( \sum_{s=1}^{N_s} {\bf d}_s - {\bf R} \right) b^2 e^{ -\beta U_{\rm sp} (z_{\rm top}) -\beta H_{\rm ext} -\beta H_{\rm self, b} -\beta H_{\rm self, s} -\beta H_{\rm bs}} \nonumber \\
&=& \int d\Gamma_b e^{-\beta H_1}  \int d\Gamma_s e^{-\beta H_2}
\, \,\frac{\int d\Gamma_b \int d\Gamma_s e^{-\beta (H_1+H_2)}e^{-\beta H_{\rm bs}}}{\int d\Gamma_b \int d\Gamma_s e^{-\beta (H_1+H_2)}} \nonumber\\
&=& e^{-\beta F_{\rm b}(N, z_{\rm top})} e^{-\beta F_{\rm s}(N_s, R)} \left<e^{-\beta H_{\rm bs}} \right>_{N,z_{\rm top}, R} \nonumber\\
&\approx & e^{-\beta F_{\rm b}(N, z_{\rm top})} e^{-\beta F_{\rm s}(N_s, R)}
e^{-\beta  \left< H_{\rm bs} \right>_{N,z_{\rm top}, R}} \nonumber
\end{eqnarray}
\end{widetext}
which yields Eq.~(\ref{eq:F_tot}) of the main text:
$$
F(N,\!z_{\rm top}, \!R)\!\approx\! F_{\rm b}(N, \!z_{\rm top}) \! +\! F_{\rm s}(N_s,\! R)\!+\!F_{\rm bs}(N,\!z_{\rm top}, R).
$$
Here $F_{\rm bs}(N,z_{\rm top}, R)= \left< H_{\rm bs} \right>_{N,z_{\rm top}, R}$. In Eq.~(\ref{eq:FNZR1}) we introduce the short-hand notations,
\begin{eqnarray}
&&\int d\Gamma_b \!= \!\int_0^{2 \pi} \!\!d\psi_1\ldots \!\!\int_0^{2 \pi} \!\!d\psi_N \int_0^{1} \!\! d \cos \theta_1 \ldots \!\! \int_0^{1} \!\! d \cos \theta_N \nonumber \\
&& \int  d\Gamma_s \! = \! \int_0^{2 \pi} d\phi_1\ldots  d\phi_{N_s} \nonumber
\end{eqnarray}
as well as
\begin{eqnarray}
&& e^{-\beta H_1} \!=\! e^{-\beta U_{\rm sp} (z_{\rm top}) -\beta H_{\rm ext} -\beta H_{\rm self, b} } \delta \!\! \left(\!z_{\rm top} - b \sum_{s=1}^N \cos \theta_s \!\!\right)b \nonumber \\
&&e^{-\beta H_2} \!=\! e^{-\beta H_{\rm self, s}} \delta \left( \sum_{s=1}^{N_s} {\bf d}_s - {\bf R} \right)b^2. \nonumber
\end{eqnarray}

To compute $\left< H_{\rm bs} \right>_{N,z_{\rm top}, R}$ we use as previously the approximation of small transverse fluctuations for the bulk part of the chain, $H_{\rm bs} \approx \left< H_{\rm bs} \right>_{\psi}$. With this approximation one can write,
\begin{eqnarray}
\label{eq:esp_bs1}
&&\left< \! e^{i {\bf k} \cdot \sum_{s=l}^N {\bf b}_{s} +i {\bf k} \cdot
\sum_{s=1}^m {\bf d}_{s}} \!\right>_{\!\!N,z_{\rm top},R}  \\
&&~~~\approx \!\! \left<  \!e^{i {\bf k}_{\perp}
\cdot \sum_{s=l}^N {\bf b}_{s}^{\perp} }\right>_{\!\!\psi}
\!\! \left< \!e^{i k_z b \cdot \sum_{s=l}^N \eta_{s} +i {\bf k}_{\perp}
\cdot \sum_{s=1}^m {\bf d}_{s}} \! \right>_{\!\!N,z_{\rm top},R} , \nonumber
\end{eqnarray}
with the same notations as above.  The first factor in the right-hand side of
Eq.~(\ref{eq:esp_bs1}) may be computed as in Eq.~(\ref{eq:Angav}),
yielding
$$
\left< e^{i {\bf k}_{\perp} \cdot \sum_{s=l}^N {\bf b}_{s}^{\perp} }\right>_{\psi} =
e^{-\frac{k_{\perp}^2b^2(N-l)}{4}\left(1-\frac{\tilde{z}_{\rm top}^2}{N^2}\right) }
=e^{-k_{\perp}^2 h_1^2}.
$$
Using the same approximation as in Eqs.~(\ref{eq:app2}) and (\ref{eq:dsR}),
$$
b\sum_{s=l}^{N}\eta_{s}\approx \frac{z_{\rm top}}{N}(N-l) = g_1;
\qquad
\sum_{s=1}^m {\bf d}_{s} =  \frac{m}{N_s} {\bf R} ={\bf R}^{\prime}
$$
we arrive at Eq.~(\ref{eq:esp_bs}), which we write as
\begin{eqnarray}
\label{eq:esp_bs2}
&&\left< \! e^{i {\bf k} \cdot \sum_{s=l}^N {\bf b}_{s} +i {\bf k} \cdot
\sum_{s=1}^m {\bf d}_{s}} \!\right>_{\!\!N,z_{\rm top},R} = e^{-k_{\perp}^2 h_1^2 +ik_z g_1 +i{\bf k}_{\perp} \cdot {\bf R}^{\prime}}
 , \nonumber
\end{eqnarray}
where $h_1$, $g_1$ and ${\bf R}^{\prime}$ have been defined in the above equations.

Below we give the calculation detail of Eq.~(\ref{eq:fbsfin}) where we need to compute the
integral in Eq.~(\ref{eq:Hsb1}) with the substitute from (\ref{eq:esp_bs}). With the above
notations for $h_1$, $g_1$ and ${\bf R}^{\prime}$ it may be written as
$$
\frac{1}{(2 \pi)^3} \int  \frac{4 \pi }{k_{\perp}^2 +k_z^2}
e^{-k_{\perp}^2 h_1^2 +ik_z g_1 +i{\bf k}_{\perp} \cdot {\bf R}^{\prime}} d {\bf k}.
$$
First we compute the integral over $k_z$ using the  residue at $k_z =i k_{\perp}$:
\begin{equation}
\label{eq:overkz}
\int_{-\infty}^{\infty}  \frac{4 \pi}{k_{\perp}^2 +k_z^2} e^{i k_z g_1 }\, dk_z
=\frac{4 \pi^2}{k_{\perp}}e^{-k_{\perp} g_1 } .
\end{equation}
Next  the integration over ${\bf k}_{\perp}$ may be performed to yield:
\begin{eqnarray}
\label{eq:overkper}
&& \frac{4 \pi^2}{8\pi^3} \int_0^{\infty} k_{\perp} dk_{\perp} \frac{e^{-k_{\perp}^2 h_1^2 -k_{\perp}g_1}}{k_{\perp}} \int_0^{2 \pi} e^{ik_{\perp} R^{\prime} \cos \phi} d\phi   \nonumber \\
&&~~~~~=\int_0^{\infty} e^{-k_{\perp}^2h_1^2- k_{\perp}g_1} J_0(k_{\perp} R^{\prime}) dk_{\perp} \\
&&~~~~~=\frac{1}{R^{\prime}} \int_0^{\infty} e^{-z^2 (h_1/R^{\prime})^2- z (g_1/R^{\prime})} J_0(z) dz \nonumber \\
&&~~~~~ \simeq \frac{1}{\sqrt{R^{\prime\, 2} +g_1^2}} \nonumber
\end{eqnarray}
where we take into account that  $g_1/R^{\prime} \gg h_1/R^{\prime}$ for $N\sim N_s \gg 1$.

Using the above result for the integral over ${\bf k}$ we can find
$\left< H_{\rm bs} \right>_{N,z_{\rm top}, R}$:
\begin{eqnarray}
\label{eq:Hbsfin1}
&&\beta \left< H_{\rm bs} \right>_{N,z_{\rm top}, R}
\simeq l_B\int_1^{N} \!\!\!dl \int_1^{N_s} \!\!\!
\frac{dm}{ \sqrt{ \frac{z_{\rm top}^2(N-l)^2}{N^2} +R^2\frac{m^2}{N_s^2}}}  \nonumber \\
&&~~ \!=  \frac{l_B NN_s}{z_{\rm top} } \!\log Z_1 \!+\!
\frac{l_BNN_s}{R}\! \log Z_2 \!+\!
\frac{l_BN}{z_{\rm top}} \!\log Z_3
\end{eqnarray}
where
\begin{eqnarray}
\label{eq:Z1Z3}
Z_1&=& (z_{\rm top}/R) + \sqrt{ 1 + (z_{\rm top}/R)^2} \\
Z_2&=& (R/z_{\rm top}) \left( 1+ \sqrt{ 1 + (z_{\rm top}/R)^2} \right) \\
Z_3&=& \frac{R}{2 z_{\rm top} N_s}
\end{eqnarray}
and we use definitions of $g_1$ and $R^{\prime}$ and approximate the summation over $l$ and $m$ by the integration. After a simple algebra we arrive at the expression (\ref{eq:fbsfin}) for $\left< H_{\rm bs} \right>_{N,z_{\rm top}, R}$.

\begin{acknowledgments}
This work was supported by a grant from the President of the RF (No MK-2823.2015.3).
\end{acknowledgments}


\end{document}